\documentclass[12pt,preprint]{aastex}
\usepackage{graphicx}
\usepackage{amssymb}
\usepackage{natbib}
\newcommand{\beq}{\begin{equation}}
\newcommand{\eeq}{\end{equation}}

\newcommand{\HST}{{\sl HST~}}

\def\fdg{\hbox{$.\!\!^\circ$}}

\begin{document}
\bibliographystyle{my2}

\title{Astrometry with the Hubble Space Telescope:
Trigonometric Parallaxes of Planetary Nebula Nuclei: NGC\,6853, NGC\,7293, Abell\,31, and DeHt\,5\footnote{Based on 
observations made with
the NASA/ESA Hubble Space Telescope, obtained at the Space Telescope
Science Institute, which is operated by the
Association of Universities for Research in Astronomy, Inc., under NASA
contract NAS5-26555}}

\author{ G.\ Fritz Benedict\altaffilmark{2}, Barbara E.
McArthur\altaffilmark{2},  Ralf Napiwotzki\altaffilmark{3}, Thomas E. Harrison\altaffilmark{4},  Hugh C. Harris\altaffilmark{5},  Edmund Nelan\altaffilmark{6}, Howard E. Bond\altaffilmark{6}, Richard J. Patterson\altaffilmark{7},   and Robin Ciardullo\altaffilmark{8} }

\altaffiltext{2}{McDonald Observatory, University of Texas, Austin, TX 78712}
\altaffiltext{3}{Centre for Astrophysics Research, STRI, University of Hertfordshire, College Lane, Hatfield AL10 9AB, UK}\altaffiltext{4}{Department of Astronomy, New Mexico State University, Las Cruces, New Mexico 88003}
\altaffiltext{5}{ United States Naval Observatory, Flagstaff Station, Flagstaff, AZ 86001}
\altaffiltext{6}{ Space Telescope Science Institute, 3700 San Martin Dr, Baltimore, MD 21218 }
\altaffiltext{7}{ Department of Astronomy, University of Virginia, PO Box 3818, Charlottesville, VA 22903}
\altaffiltext{8}{ Department of Astronomy and Astrophysics, The Pennsylvania State University, University Park, PA 16802}



\begin{abstract}
We present absolute parallaxes and relative proper motions for the central stars of the planetary nebulae  NGC\,6853 (The Dumbbell), NGC 7293 (The Helix), Abell 31,  and  DeHt 5.  This paper details our reduction and analysis using DeHt\,5 as an example. We obtain these planetary nebula nuclei (PNNi) parallaxes with astrometric data from Fine Guidance Sensors
FGS 1R and FGS 3, white-light interferometers on the Hubble Space Telescope ({\it HST}). Proper motions, spectral classifications and VJHKT$_2$M and DDO51 photometry of the stars comprising the astrometric reference frames provide spectrophotometric estimates of reference star absolute parallaxes. Introducing these into our model as observations with error, we determine absolute parallaxes for each PNN. Weighted averaging with previous independent parallax measurements yields an average parallax precision, $\sigma_{\pi}/\pi =  5$\%. Derived distances are: d$_{NGC\,6853}=405^{+28}_{-25}$pc, d$_{NGC\,7293}=216^{+14}_{-12}$pc,  d$_{Abell\,31} =  621^{+91}_{-70}$pc, and d$_{DeHt\,5} = 345^{+19}_{-17}$pc. These PNNi distances are all smaller than previously derived from spectroscopic analyses of the central stars. To obtain absolute magnitudes from these distances requires estimates of interstellar extinction. We average extinction measurements culled from the literature, from reddening based on PNNi intrinsic colors derived from model SEDs, and an assumption that each PNN experiences the same rate of extinction as a function of distance as do the reference stars nearest (in angular separation) to each central star. We also apply Lutz-Kelker bias corrections. The absolute magnitudes and effective temperatures permit estimates of PNNi radii,  through both the Stefan-Boltzmann relation and Eddington fluxes. Comparing absolute magnitudes with post-AGB models provides mass estimates. Masses cluster around 0.57 {\cal M}$_{\sun}$, close to the peak of the white dwarf mass distribution.
~Adding a few more PNNi with well-determined distances and masses, we compare all the PNNi  with cooler white dwarfs of similar mass, and confirm, as expected, that PNNi have larger radii than white dwarfs that
have reached their final cooling tracks.

\end{abstract}




%

\section{Introduction}

Planetary nebulae (PNe) are a visually spectacular and relatively short-lived step in the evolution from asymptotic giant branch (AGB) stars to a final white dwarf (WD) stage. \cite{Ibe83} first argued that the ejection of most of the gaseous envelope in AGB stars occurs during the thermal pulse phase, in the form of a massive, low-velocity wind. As summarized by
\cite{Sta02},  the remnant central star (PN nucleus, PNN; PN nuclei, PNNi) ionizes the gaseous ejecta, while a fast, low mass-loss rate PNN wind shapes the PN. PN morphology depends on a complicated combination of phenomena, some occurring within the nebular gas, which evolves on a dynamical timescale, and others caused by multiplicity and/or the evolution of the stellar progenitors and of the PNN. Morphology may also depend on the physical status of the interstellar environment of the PN progenitor.

Intercomparison of PNe can aid our understanding of the complicated astrophysics of this stage of stellar evolution, particularly if distances are known.  Many indirect methods of PN distance determination exist (e.g., \cite{Cia99} and \cite{Nap01}), including estimates from interstellar Na D lines, NLTE stellar atmospheres analyses of the PNNi (e.g. Hultzsch et al. 2007 \nocite{Hul07}), estimates from resolved companion stars \citep{Cia99}, and from Galactic kinematics \citep{Nap06}. The expansion method (e.g. Palen et al. 2002 \nocite{Pal02}) becomes model-dependent when applied to PNe with asymmetric or irregular geometry, and
requires either an assumption that apparent expansion is due to material
motions, or that one models the motion of the ionization front. Agreement among these distance determination methods is seldom better than 20\%, often worse. Direct parallax measurments of PNNi rarely have precisions smaller than the measured parallax, a notable exception being \cite{Har97} and \cite{Har07}, who, using narrow-field CCD astrometry, provide $\sim$0.4 millisecond of arc (mas) precision parallaxes for  PNNi nearer than $\sim$500 pc. 

To further reduce the distance errors for a few PNNi (chosen as nearby from Harris et al. 1997), we have determined new absolute parallaxes of the PNNi of DeHt5, Abell\,31, and NGC\,7293 (The Helix), using FGS1r.  Positions and aliases are given in Table 1. We have also determined a revised parallax for NGC\,6853 (The Dumbell = M27), using previously collected FGS 3 data (Benedict et al. 2003\nocite{Ben03}), combined with several new FGS1r measurements. Our present errors average 2-3 times smaller than those in Harris et al (2007). However, some reduction in the final parallax errors is obtained
through a weighted average of our present parallaxes with those in \cite{Har07}. \cite{Nap95} classifies all the PNNi considered in this paper as WD of type DAO or DA. 

Our reduction and analysis of these data is basically the same for our previous work on NGC\,6853 (Benedict et al. 2003). Our extensive investigation of the astrometric reference stars provides an independent estimation of the line of sight extinction as a function of distance for these PNNi, a significant contributor to the uncertainty in the absolute magnitude, M$_V$. Using DeHt\,5 as an example throughout, we present the results of extensive spectrophotometry of the astrometric reference stars, information required to derive absolute parallaxes from relative measurements; briefly discuss data acquisition and analysis; extract limits  on binarity and photometric variability; and derive an absolute parallax for each PNNi. Finally, from a weighted average of our new parallaxes and those of Harris et al. (2007) we calculate an absolute magnitude for each PNN and derive stellar radii. With these and estimates of PNN mass  from post-AGB evolution models we derive surface gravities, log $g$, to compare with those from the
\cite{Nap99} stellar atmosphere analyses. We discuss some astrophysical consequences of these new, more precise distances and summarize our findings in Section~\ref{SUMRY}.

\cite{Bra91} and \cite{Nel07}
provide an overview of the
FGS instrument and \cite{Ben99}, \cite{Ben02a}, \cite{Har04},  \cite{Ben07} describe the fringe tracking (POS) mode astrometric capabilities 
of an FGS, along with the data acquisition and reduction strategies used in the present study. 
We time-tag all data with a modified Julian Date, $mJD = JD - 2400000.5$.

\section{Observations and Data Reduction}  \label{AstRefs}

Using DeHt\,5 as an example, Figure \ref{fig-1} shows the distribution on the sky of the ten reference stars and the PNN. This image is from the Digitized Sky Survey, via {\it Aladin}. For this target fifteen sets of astrometric data were acquired with FGS 1r on {\it HST}, spanning 3.8 years, for a total of 248 measurements of the DeHt\,5 PNN  and reference stars. 
Each data set required approximately 33 minutes of spacecraft time. The data were reduced and calibrated as detailed in \cite{Ben02a},  \cite{Ben02b}, \cite{mca01}, and \cite{Ben07}. At each epoch we measured reference stars and the target  multiple times, this to correct for intra-orbit drift of the type seen in the cross filter calibration data shown in figure 1 of \cite{Ben02a}. 

Table~\ref{tbl-LOO} lists the fifteen epochs of observation and highlights another particular difficulty with these data. Ideally ({\it cf.} Benedict et al. 2007) we obtain observations at each of the two maximum parallax factors\footnote{Parallax factors are projections along RA and Dec of the Earth's orbit about the barycenter of the Solar System, normalized to unity.} at two distinct spacecraft roll values imposed by the requirement that
{\it HST} roll to provide thermal control of the camera in the radial bay and to keep its solar panels fully illuminated throughout
the year.  This roll constraint generally imposes alternate orientations at each time of maximum positive or negative parallax factor over a typical two year campaign. A few observations at intermediate or low parallax factors usually allows a clean separation of parallax and proper motion signatures. In the case of DeHt\,5, (as well as Abell\,31 and NGC\,7293), two-gyro guiding\footnote{\HST has a full compliment of six rate gyros, two per axis, that provide coarse pointing control. By the time these observations were in progress, three of the gyros had failed. \HST can point with only two. To ``bank" a gyro in anticipation of a future failure, NASA decided to go to two gyro pointing as standard operating procedure.} forced us into the less than satisfactory distribution of parallax factors shown in Table~\ref{tbl-LOO}.  Specifically for DeHt 5, there are no large positive parallax factors in Right Ascension. However, the higher the absolute declination of the target, the more likely it is that there will be windows of visibility near times of $\pm$ maximum parallax factor, either in RA or declination. Additionally, large declination typically results in higher ecliptic latitude.  The ecliptic latitude of DeHt\,5 renders its parallactic ellipse rather round, increasing the value of observations  that were forced to be secured at times far from maximum parallax factor.  We gain parallactic displacement in declination at the expense of displacement in RA.
These scheduling and solar-panel illumination constraints also resulted in the spotty access to our reference stars indicated in Table~\ref{tbl-LOO}. 

Finally, for DeHt\,5 and Abell\,31 we were able to take advantage of science instrument command and data handling (SIC\&DH) computer problems that took  the only other then operational science instrument (WFPC2) off-line in late 2008. This situation opened a floodgate  of FGS proposals, temporarily rendering \HST nearly an 'all astrometry, all the time' mission. Consequently, we obtained another epoch well-separated in time from the original eleven. This permitted a significantly better determination of relative proper motion for these two targets.

\section{PNNi Photometry and Companion Limits} 

The FGS have two operating modes. We used both. The fringe tracking (POS mode) data collected over the course of this project can be used to establish the degree of  photometric variability of these PNNi. Additionally, fringe scanning (TRANS mode) data provide an opportunity to discover previously unknown companions, or to establish limits for separation and magnitude difference.

\subsection{FGS Photometry of the PNNi} \label{PHOT}
FGS 1r and FGS 3 are precision photometers, yielding relative photometry with internal error  $\gtrsim 0.002$ mag (Benedict et al. 1998a\nocite{Ben98a}). During each of the observation sets we observed the  PNNi 4--7 times over approximately 33 minutes. Given the faintness of these stars, we made no effort to explore for high-frequecy variations during a single observation.\footnote{The FGS samples the fringe zero crossing at a 40Hz rate. See \cite{Ben98b} for an example of the use of an FGS as a high-speed photometer, capturing a flare event in the vicinity of Proxima Cen.} Standard deviations within any one observation sets were on order 0.1\% for the two brighter PNNi (NGC\,6853 and NGC\,7293), 0.2\% for Abell\,31, and 0.5\% for DeHt\,5. We derived a PNNi average intensity for each observation set and used the summed intensity of the brightest astrometric reference stars as a flat field. The resulting intensities were then transformed to relative magnitudes such that the average relative magnitude over the entire campaign matched the measured magnitude from \cite{Har07}. 

We show a montage of the resulting PNNi light curves in Figure~\ref{FigLC}. Our coverage is too sparse to extract any believable periodic astrophysical component to these variations. However, the internal errors for three PNNi suggest variations far larger than those expected from photometric errors only. This was expected. We fit  these variations with sinusoids with periods of one year, because of previously encountered position dependent variations of the brightnesses of the stars used to generate the flat field (see Benedict et al. 1998a, section 2.3.1 for a discussion of roll-induced photometric variation). All PNNi were observed at similar positions within the FGS, typically with radial distances from pickle center $\le$10\arcsec. Variations introduced by Zodiacal light will be insignificant for the two brightest PNNi and for DeHt\,5 at a high ecliptic latitude. Abell\,31 is closest to the ecliptic, but nearly all observations were obtained at essentially identical Sun-target angular separations, resulting in little variation due to changing background. 
From  Figure~\ref{FigLC} and the statistics of intra-orbit observations we conclude that an upper limit to photometric 'flickering' on timescales of minutes to years is around 5 mmag.

\subsection{Assessing PNNi Binarity} \label{BINLIM}
Companions of stellar and substellar mass have been invoked to produce the asymmetric structure seen in PN \citep{Bon90,Sok06,DeM09}. Using FGS 1r TRANS mode observations (e.g., Franz et al. 1998 \nocite{Fra98}, 
Nelan et al. 2004 \nocite{Nel04}), we have analyzed the fringe morphology
of these PNNi and find that all are unresolved. 
This places limits on separation and magnitude difference, $\Delta m$,
for possible companions. Details on detectability can be found in Section 3.3.2
of the FGS Instrument Handbook (Nelan 2007). Summarizing the complex
interplay among system magnitude, $\Delta m$, and separation, we would have
detected companions with
separations of 10 mas and larger with $\Delta m \le 1$. Detectability at separations of 15 mas
increases to $\Delta m \le 2$. For separations $\ge 50$ mas FGS 1r achieves a detectability of
$\Delta m \le 3.5$.
Figure~\ref{Fig-SC} compares the X and Y axis fringes of DeHt\,5 with those from Abell\,31.
Either the two objects have identical companions, 
or they are both unresolved. Similar comparisons were made 
between NGC\,6853 and NGC\,7293 with similar results. Once we determine absolute magnitudes, parallaxes, and estimate masses for these PNNi  we will (Section~\ref{BINLIM2} below) establish the spectral types, separations in AU, and periods for companions that would remain hidden from the FGS.

\section{Spectrophotometric Absolute Parallaxes of the Astrometric Reference Stars} \label{SpecPhot}
Because the parallax determined for the PNNi will be
measured with respect to reference frame stars which have their own
parallaxes, we must either apply a statistically derived correction from relative to absolute parallax (Van Altena, Lee \& Hoffleit 1995, hereafter YPC95 \nocite{WvA95}) or estimate the absolute parallaxes of the reference frame stars listed in Table \ref{tbl-LOO}. In principle, the colors, spectral type, and luminosity class of a star can be used to estimate the absolute magnitude, M$_V$, and V-band absorption, A$_V$. The absolute parallax is then simply,
\beq
\pi_{abs} = 10^{-(V-M_V+5-A_V)/5}
\eeq
The luminosity class is generally more difficult to estimate than the spectral type (temperature class). However, the derived absolute magnitudes are critically dependent on the luminosity class. As a consequence we obtained additional photometry in an attempt to confirm the luminosity classes. Specifically, we employ the technique used by \cite{Maj00} to discriminate between giants and dwarfs for stars later than $\sim$ G5, an approach also discussed by  \cite{Pal94}.

\subsection{Broadband Photometry}
Our band passes for reference star photometry include:  BV (CCD photometry from a 1m telescope at New Mexico State University) and JHK (from 2MASS\footnote{The Two Micron All Sky Survey
is a joint project of the University of Massachusetts and the Infrared Processing
and Analysis Center/California Institute of Technology }). We also had access to Washington/DDO filters M, T$_2$, and DDO51 (obtained at Fan Mountain Observatory with the 1m, and at Las Campanas Observatory with the Swope 1m).  
Table \ref{tbl-SPP} lists the visible and infrared photometry for the DeHt\,5  reference stars, ref-2 through ref-11.

\subsection{Spectroscopy, Luminosity Class-sensitive Photometry, and Reduced Proper Motion}
The spectra from which we estimated spectral type and luminosity class come from the New Mexico State University Apache 
Point Observatory\footnote{ The
Apache Point Observatory 3.5 m telescope is owned and operated by
the Astrophysical Research Consortium.}. The dispersion 
was 0.61 \AA/pixel with wavelength coverage 4101 -- 4905 \AA, yielding R$\sim$3700. Classifications used a combination of template matching and line ratios. The brightest targets had about 1500 counts
above sky per pixel, or S/N $\sim$ 40, while the faintest targets had about 400 counts
per pixel (S/N $\sim$ 20). The spectral types for the higher S/N stars are within $\pm$1 subclass. Classifications for the lower S/N stars are $\pm$2
subclasses. Table \ref{tbl-SPP} lists the spectral types and luminosity classes for our reference stars. 

The Washington/DDO photometry  can provide a possible confirmation of the estimated luminosity class, depending on the spectral type and luminosity class of the star (later than G5 for dwarfs, later than G0 for giants). Washington/DDO photometry was more helpful as a discriminator for the DeHt\,5 field than for our previous work on NGC\,6853 (e.g., Benedict et al. 2003\nocite{Ben03}), suggesting a giant luminosity classification for ref-9 and ref-11. However, the problems related to the NGC\,6853 PN nebulosity  discussed in that paper also affected the NGC\,7293 reference frame DDO photometry. 

We  employ the technique of reduced proper motions to provide a confirmation of the reference star estimated luminosity class listed in Table~\ref{tbl-SPP}. We obtain preliminary proper motions ($\mu$)  from UCAC2 \citep{Zac04} and $J$, $K$ photometry from 2MASS for a one-degree-square field centered on DeHt 5. With final proper motions from our astrometric solution (Section~\ref{AST}) we plot Figure~\ref{fig-RPM}, which shows $H_K = K + 5\log(\mu)$ versus $(J-K)$ color index for 458 stars. If all stars had the same transverse velocities, Figure~\ref{fig-RPM} would be equivalent to an H-R diagram. DeHt 5 and reference stars are plotted as ID numbers from Table~\ref{tbl-SPP}. Errors in $H_K$ are now $\sim0.3$ mag. Reference stars 9, and 11 are clearly separated from the others, supporting their classification as giants. 

\subsection{Interstellar Extinction} \label{AV}
To determine interstellar extinction we first plot these stars on several color-color diagrams. A comparison of the relationships between spectral type and intrinsic color against those we measured provides an estimate of reddening. Figure \ref{fig-CCD} contains  a J-K vs V-K color-color diagram and reddening vector for A$_V$ = 1.0. Also plotted are mappings between spectral type and luminosity class V and III from \cite{Bes88} and \cite{Cox00} (hereafter AQ2000). Figure~\ref{fig-CCD}, and similar plots for the other measured colors, along with the estimated spectral types, provides an indication of the reddening for each reference star. 

Assuming an R = 3.1 galactic reddening law (Savage \& Mathis 1979\nocite{Sav79}), we derive A$_V$ values by comparing the measured colors (Table~\ref{tbl-SPP} ) with intrinsic B-V, J-K, and V-K colors from \cite{Bes88} and AQ2000. Specifically we estimate A$_V$ from three different ratios, each derived from the Savage \& Mathis (1977) reddening law: A$_V$/E(J-K) = 5.8; A$_V$/E(V-K) = 1.1; and A$_V$/E(B-V) = 3.1.  The resulting average A$_V$ are collected in Table \ref{tbl-SPP}. 

\subsection{Adopted Reference Frame Absolute Parallaxes}

We derive absolute parallaxes with M$_V$ values from AQ2000 and the $\langle$A$_V\rangle$ derived from the photometry. Our parallax values are listed in Table \ref{tbl-SPP}. We produce errors on the absolute parallaxes by combining contributions from uncertainties in M$_V$ and A$_V$, which we have combined and set to 0.5 magnitude for each reference star. Individually, no reference star parallax is better determined than ${\sigma_{\pi}\over \pi}$ = 23\%.  The average absolute parallax for the DeHt 5 reference frame is $\langle\pi_{abs}\rangle = 1.0$ mas.
As a sanity check we compare this to the correction to absolute parallax discussed and presented
in YPC95 (section 3.2, fig. 2). Entering
YPC95, fig. 2, with the DeHt 5 galactic
latitude, l = -12\arcdeg, and average magnitude for the
reference frame, $\langle V_{ref} \rangle$= 14.3, we obtain a galactic model-dependent correction
to absolute of 1.0 mas, in agreement.

\section{Absolute Parallaxes of the PN Central Stars}
Sections 5.1-5.4 detail our astrometric modeling of the DeHt\,5 data. Any differences in modeling for other PNNi are noted in Section~\ref{ASTnotes}, below. We compare our new distances with other more indirect estimates later in Section~\ref{Dist}.

\subsection{The DeHt\,5 Astrometric Model}\label{AST}

 With the positions measured by FGS 1r we determine the scale, rotation, and offset ``plate
constants" relative to an arbitrarily adopted constraint epoch (the so-called ``master plate") for
each observation set (the data acquired at each epoch). The mJD of each observation set is listed in Table~\ref{tbl-LOO}.
The DeHt\,5 reference frame contains 10 stars. We employ an eight parameter model for those observations. 
For the DeHt\,5 field all the reference stars are redder than the science target. Hence, we also apply the corrections for lateral color discussed in Benedict et al. (1999). 

As for all our previous astrometric analyses, we employ GaussFit \citep{Jef88} to minimize $\chi^2$. The solved equations
of condition for DeHt\,5 are:
\beq
        x' = x + lc_x(\it B-V) 
\eeq
\beq
        y' = y + lc_y(\it B-V) 
\eeq
\beq
\xi = Ax' + By' + C + R_x (x'^2 + y'^2) - \mu_x \Delta t  - P_\alpha\pi_x
\eeq
\beq
\eta = Dx' + Ey' + F + R_y(x'^2 + y'^2) - \mu_y \Delta t  - P_\delta\pi_y
\eeq
where $\it x$ and $\it y$ are the measured coordinates from {\it HST};
$\it lcx$ and $\it lcy$ are the
lateral color corrections from Benedict et al. 1999\nocite{Ben99}; and $\it B-V $ are
those  colors for each star. A,  B, D, and E   
are scale and rotation plate constants, C and F are
offsets; $R_x$ and $R_y$ are radial term coefficients;
$\mu_x$ and $\mu_y$ are proper motions; $\Delta$t is the epoch difference from the mean epoch;
$P_\alpha$ and $P_\delta$ are parallax factors;  and $\it \pi_x$ and $\it \pi_y$
 are  the parallaxes in x and y. We obtain the parallax factors (projections along RA and Dec of the Earth's orbit about the barycenter of the Solar System normalized to unity) from a JPL Earth orbit predictor (\cite{Sta90}), upgraded to version DE405. 
 
\subsection{Prior Knowledge and Modeling Constraints} \label{MODCON}
In a quasi-Bayesian approach the reference star spectrophotometric absolute parallaxes (Table~\ref{tbl-SPP}) and proper motion estimates for DeHt\,5 (Harris et al.  2007) and for the reference stars  from UCAC2 (Zacharias et al. 2004)  were input as observations with associated errors, not as hardwired quantities known to infinite precision.  Input proper motion values  have typical errors of 4--6 mas y$^{-1}$ for each coordinate.  The lateral color calibration and the B-V color indices are also treated as observations with error.  Orientation to the sky is obtained from ground-based astrometry 
from 2MASS with uncertainties in the field orientation $\pm 0\fdg05$. This value, too, was made available as an observation with error. We essentially model a 3$^D$ volume of the space that contains our science target and reference stars, all at differing distances.

\subsection{Assessing Reference Frame Residuals}
The Optical Field Angle Distortion calibration (McArthur et al. 2002\nocite{McA02}) reduces as-built {\it HST} telescope and FGS 1r distortions with amplitude $\sim1\arcsec$ to below 2 mas over much of the FGS 1r field of regard. From histograms of the reference star astrometric residuals (Figure~\ref{fig-4}) we conclude that we have obtained satisfactory correction in the
region available at all {\it HST} rolls. The resulting reference frame 'catalog' in $\xi$ and $\eta$ standard coordinates (Table \ref{tbl-POS}) was determined
with	$<\sigma_\xi>= 0.3$	 and	$<\sigma_\eta> = 0.3$ mas. Relative proper motions along RA (x) and Dec (y) are also listed in Table \ref{tbl-POS}. The proper motion vector is listed in Table~\ref{tbl-SUM}, as are astrometric results for the other PNNi, including catalog statistics.

To determine if there might be unmodeled - but possibly correctable -  systematic effects at the 1 mas level, we plotted the DeHt\,5 reference frame x and y residuals against a number of spacecraft, instrumental, and astronomical parameters. These included x,y  position within the pickle-shaped FGS field of regard; radial distance from the center of the FGS field of regard; reference star V magnitude and B-V color; and epoch of observation.  We saw no obvious trends, other than an expected increase in positional uncertainty with reference star magnitude. 

\subsection{The Absolute Parallax of the DeHt\,5 Central Star} \label{AbsPi}
For this object at high ecliptic latitude (note the large parallax factors in both RA and Dec in Table~\ref{tbl-LOO}) we can solve for the separate x and y components of the parallax. These were $\pi_x=2.79 \pm 0.17$ mas and $\pi_y = 3.19 \pm 0.35$ mas. We obtain for the DeHt\,5 PNN a final absolute parallax $\pi_{abs} = 2.86 \pm0.16$ mas. Our result agrees within the errors with the previous ground-based parallax measurement of the DeHt\,5 PNN (Harris et al. 2007), $\pi_{abs} = 3.34 \pm0.56$ mas. 
Parallaxes from {\it HST} and {\it USNO} and relative proper motion results from {\it HST} are collected in Table~\ref{tbl-SUM}. Even though both proper motion determinations are relative, using different sets of reference stars, the proper motion vector agrees  with that determined by USNO (Harris et al. 2007). For the remainder of this paper we adopt as the absolute parallax of the DeHt\,5 PNN, $\pi_{abs} = 2.90 \pm0.15$ mas, the weighted average of these two independent parallax determinations.  The degree of independence is quantified in Table~\ref{tbl-SUM}. Of the ten reference stars used in the \HST study, only three were in the suite of fifteen (on average much fainter) reference stars used in the {\it USNO} study. If
both studies used exactly the same set of reference stars, one would expect that
some component of the uncertainties would be correlated. As indicated in Table~\ref{tbl-SUM}, this is not the case for any of the PNNi investigated in our study.

\subsection{Modeling Notes on the Other PNNi}\label{ASTnotes}

{\bf Abell\,31} - This field provided six reference stars. The reference star average data are listed in Table~\ref{tbl-SUM}. We again used the eight parameter model (Equations 4 and 5). Because of the low ecliptic latitude, most of the parallax signature is along RA. Hence, we constrained $\pi_x = \pi_y$. Two gyro guiding scheduling constraints and the afore-mentioned science-side problems yielded a total study duration of four years. The \HST parallax, $\pi_{abs} = 1.51 \pm0.26$ mas agreed well with the USNO value from Harris et al. (2007),  $\pi_{abs} = 1.76 \pm0.33$.  For the remainder of this paper we adopt as the absolute parallax of the Abell\,31 PNN, $\pi_{abs} = 1.61 \pm0.21$ mas, the weighted average of these two  independent parallax determinations. \HST, USNO,  and final \HST + USNO weighted average parallaxes are given in Table~\ref{tbl-SUM}. This distance, d= $621^{+91}_{-70}$ pc, rules out a physical association between DeHt\,5 and the red companion detected by the \HST WFPC-2 camera discussed in \cite{Cia99}.

{\bf NGC\,7293} -  This field provided only three useful reference stars. The reference star average data are listed in Table~\ref{tbl-SUM}. Because of the paucity of reference stars, the astrometric model for this field used only four parameters, discarding the radial terms and constraining d=-b, e=a in Equations 4 and 5. We also constrained $\pi_x = \pi_y$. Model selection was  dictated by the loss of access to 1-2 reference stars for many of the observation sets, primarily due to two-gyro guiding constraints on allowed spacecraft roll. One of the reference stars (\#18 in the original GO-10432 proposal) was removed from our modeling. For several observation sets each visit to that reference star locked on a different component. From these systematic residuals after initial modeling we inferred that it is either a binary or an optical double with component separation of $\sim11$ mas.    
Our NGC\,7293 PNN parallax is $\pi_{abs} = 4.67 \pm0.16$ mas. 

\cite{Lee07a} asserts that the intrinsic width of the main sequence for FGK stars is 0.4 mag (1-$\sigma$). Past results \citep{Ben07} indicate that for fields with 5 or more reference stars, cosmic dispersion in reference star absolute magnitude has no apparent consequence. In fact, Cepheid astrophysics argues that our parallax errors are overstated. To obtain a unity reduced
$\chi^2$ ($\chi^2$ / DOF where DOF = degrees of freedom) for
our linear M$_V$ - logP Period-Luminosity relation, we must significantly
reduce our magnitude errors. For this one target whose parallax is dependent on only three reference stars we explored the effects of cosmic dispersion on reference star input absolute parallaxes. Worst-case (1-$\sigma$ increase or decrease in the M$_V$ of all reference stars) the final absolute parallax for NGC 7293 could range $\pm0.29$ mas. Hence we add in quadrature that error due to cosmic dispersion to our astrometry-only error for a final result of  $\pi_{abs} = 4.67 \pm0.33$ mas. \cite{Har07} obtained $\pi_{abs} = 4.56 \pm0.49$ mas.  A weighted average yields $\pi_{abs} = 4.66 \pm0.27$ mas.  Our measured proper motion vector does not agree within the errors with the USNO value. This is not unexpected, given that our proper motion is measured against so few reference stars.

{\bf NGC\,6853} - As noted in \cite{Ben03} our original FGS 3-only data did not adequately cover the epochs of maximum parallax factor, resulting in a relatively (for \HST) poorly determined parallax, $\pi_{abs} = 2.10 \pm0.48$ mas. The addition of two new epochs of observation at maximum parallax factor with FGS 1r significantly increased the duration of the study (now over nine years) and improved the precision and slightly increased the parallax, now $\pi_{abs} = 2.22 \pm0.19$ mas. The astrometric model for this field used only six parameters, discarding the radial terms in Equations 4 and 5, a choice dictated by an insignificant decline in reduced $\chi ^2$ when increasing from 6 to 8 astrometric coefficients (Equations 4 and 5). We also constrained $\pi_x = \pi_y$. For this object there are two USNO results (Harris et al. 1997, Harris et al. 2007) which yield the weighted USNO average $\pi_{abs} = 3.17 \pm0.32$ mas and the final \HST + USNO weighted average $\pi_{abs} = 2.47 \pm0.16$ mas listed in Table~\ref{tbl-SUM}. We note the significant difference in relative proper motion between \HST and USNO.

\section{PNNi Absolute Magnitudes, Radii, and Masses}

Again, we use DeHt 5 as an example of the steps required to obtain absolute magnitudes and radii for these PNNi. The final results for the other three PNNi are summarized in Table~\ref{tbl-AQ} and in individual notes (Section~\ref{notes-MV}) below. 
\subsection{Absolute Magnitudes and the Lutz-Kelker-Hanson Bias}

When using a trigonometric parallax to estimate the absolute
magnitude of a star, a correction should be made for the
Lutz-Kelker  bias \citep{Lut73} as modified by \cite{Han79}. See \cite{Ben07}, section 5, for a more detailed rationale for the application of this correction to single stars.
Because of the galactic latitude and distance of DeHt 5, 
and the scale height of the
stellar population of which it is a member,
we calculate Lutz-Kelker-Hanson (LKH) bias twice, assuming first a spheroidal then a disk distribution.  The LKH bias is proportional to $(\sigma_{\pi}/\pi)^2$. Presuming that the PNN belongs to the same class of object as $\delta$ Cep (young, evolved Main Sequence stars in a core helium burning phase), we scale the LKH correction determined for $\delta$ Cep in Benedict et al. (2002b) and obtain LKH = -0.02 magnitude.  Presuming that the PNN belongs to the same class of object as RR Lyr (older, evolved Main Sequence stars on the horizontal branch), we scale the LKH correction determined for RR Lyr in Benedict et al. (2002a) and obtain LKH = -0.03 magnitude. Our final LKH bias corrections are an average of the biases from the two adopted prior distributions. The corrections differ by at most 0.02 magnitude. See Benedict et al. (2007), section 5, for a more detailed rationale, justifying the use of this correction to single stars.

\subsection{The Absolute Magnitude of the PNN of DeHt\,5}
Adopting for the DeHt 5 PNN  
V= 15.47$\pm$ 0.03  and  the weighted average absolute parallax, $\pi_{abs} = 2.90 \pm0.15$ mas from Table \ref{tbl-SUM}, we determine a distance modulus, m-M = 7.69$\pm 0.12$. To obtain a final absolute magnitude, we must correct for interstellar extinction. 
There are a variety of techniques used to estimate the extinction towards
(and internal to) planetary nebulae. Perhaps the most common is the 
assumption that in nebular conditions, the ratio of H$\alpha$/H$\beta$ = 2.86,
and any deviation from this value is assigned to extinction. Alternatively,
the observed radio flux can lead to a prediction of the flux in H$\beta$
assuming an optically thin nebula with a temperature of 10$^{\rm 4}$ K. Either
technique leads to a value for a logarithmic extinction at H$\beta$ usually
denoted by ``c'', where E($B$  $-$  $V$) = 0.83c \citep{Mil75}.
Other techniques include assuming a single value for the color for the PN 
central star, using interstellar absorption features seen in the optical
or ultraviolet spectra, or simply using field stars along similar lines of 
sight.  

For DeHt\,5 two estimates of  ``c'' (\nocite{Pot96, Phi05}Pottasch
1996, Phillips 2005) list E($B$  $-$  $V$) = 0.0. \cite{Har07}
derive E($B$  $-$  $V$) = 0.18, assuming 
($B$  $-$  $V$)$_{\rm o}$ = $-$0.38, yielding (R = $A_V$/E(B-V) = 3.1) A$_V^* = 0.56$.  Estimating with field stars, from Table \ref{tbl-SPP} (Section \ref{AV}) we  derive a per-star, per-unit 100 pc distance absorption, $\langle A$$_V$$\rangle$/100pc. The average of the three stars nearest the central target (see Figure~\ref{fig-1}), ref-2, -4, and -6,  is $\langle$A$_V\rangle$/100pc = 0.09$\pm$0.01. With this per-unit 100 pc $\langle$A$_V\rangle$ and the distance to the DeHt 5 central star, d = 345$^{+19}_{-17}$ pc, we obtain a total absorption for the PNN, A$_V^* = 0.32\pm0.03$.  We also estimate A$_V$ using the measured temperature and a grid of synthetic spectra of hot, compact stars \cite{Rau03}. We calculate an intrinsic ($B$  $-$  $V$)$_{\rm o}$ = $-$0.36. This yields A$_V^* = 0.42 \pm 0.07$, where the error is obtained through a 50000 trial Monte-Carlo process. Given the scatter in A$_V$ from the various determinations, we choose to average the determination from synthetic spectra and the reference star values, yielding A$_V^* = 0.37 \pm 0.07$, deeming the 'c' determination flawed for this faint an object.  Including the LKH correction we obtain an absorption-corrected magnitude, V$_0$ = 15.03. The distance modulus and V$_0$ provide an absolute magnitude M$_V = 7.39^{+0.14}_{-0.14}$.

\subsection{A Radius for the PNN of DeHt\,5}\label{RAD}
We employ two methods to derive a radius, both differential in nature. The first method employs the Stefan-Boltzmann relation, the second filter-averaged Eddington fluxes. 

To estimate a radius, $R_{*}$, for this star  using the Stefan-Boltzmann relation we require a distance, an absolute magnitude, an effective temperature, T$^{*}_{eff}$, and a bolometric correction (B.C.) . These quantities then yield a radius via differential comparison with the 
Sun. Our parallax provides a distance, d = 345$^{+19}_{-17}$ pc and an absolute magnitude, M$_V = 7.39\pm0.14$. \cite{Nap99} has estimated T$^{*}_{eff} = 76,500 K \pm$ 5800 K from model atmosphere fits to the Balmer H$\delta$ and H$\epsilon$ absorption lines.  We calculated bolometric corrections from our synthetic photometry, by comparing the integrated surface flux $\sigma T_{\mathrm{eff}}^4$ with the surface flux through the $V$ band filter including the filter constant (or in other words computing the offset between $M_{\mathrm{bol}}$ and $M_V$ of a star with arbitrary radius). We set the zeropoint by adopting the following values for the Sun: $M_{\mathrm{bol}}^\odot = 4.75$ and $M_V^\odot = -26.74$. For DeHt\,5 
we calculate  B.C. = - 5.84 $\pm$ 0.2, where the error is dominated by the uncertainty in T$^{*}_{eff}$.

We obtain a PNN 
bolometric luminosity M$_{bol}$  = M$_V$ + B.C. = +1.55 $\pm 0.24$. $R_{*}$ follows from the expression
\beq
M_{bol}^{\sun} - M_{bol}^{*} = 10~log(T_{eff}^{*}/~T_{eff}^{\sun})+ 5~log(R_{*}/R_{\sun})
\eeq
where we assume for the Sun $T_{eff}^{\sun}=5800 K$. 
We find $R_{*} = 0.026 \pm 0.007 R_{\sun}$. The sources of error for this radius are in the absolute magnitude (i.e., the parallax), the
bolometric correction, and the $T_{eff}^{*}$. 

A second way to obtain $R_{*}$ involves the V-band average Eddington flux, $H_V$, discussed in 
\cite{Ber95} and \cite{Hol06}. The latter carried out a careful photometric calibration of DA white dwarfs with pure hydrogen atmospheres. However, radiative levitation in the hottest white dwarfs causes metals to be present in the atmospheres on a level roughly equivalent to solar abundances (Barstow et al. 2003)\nocite{Bar03}.  This has some impact on the SED causing redistribution of flux from the UV to longer wavelengths. To take this into account we computed synthetic photometry in the Johnson system from NLTE model spectra calculated with solar abundances of important elements up to the iron group \citep{Rau03}. The photometry is linked to the Vega system as outlined by Holberg \& Bergeron (2006). The overall effect of the metals is that $B-V$ colours for hot PNNi are bluer by $\approx$0.03\,mag, and the flux level in the $V$ band increases by typically 0.2\,mag compared to pure hydrogen atmospheres.

Our synthetic photometry
provides $H_V^{*}$ as a function of temperature for solar-metallicity WD of various temperatures calibrated against Vega. We obtain $H_V^{*}$ for $T^{*}_{eff}=76,500$ K. With $H_V^{Vega}$, we can derive $R_{*}$ from
\beq
R_{*}^2 = (H_V^{Vega}/H_V^{*})10^{-0.4(M_V^{*}-M_V^{Vega})}
\eeq
We obtain for DeHt\,5 with $T^{*}_{eff}=76,500 K$, $M_V^{*} = 7.39$ from our parallax, and $M_V^{Vega} = +0.026$ an $R_{*}=0.025 \pm 0.002 R_{\sun}$, where the radius error is obtained through a 50000 trial Monte-Carlo process.  Given that the approach relying directly on the B.C. and the approach utilizing $H_V$ yield $R_{*}$ values that agree, we will use for the remainder of this paper the lower error, $R_{*}=0.025 \pm 0.002 R_{\sun}$. For the other three objects the radius from Stefan-Boltzmann will be calculated as a confirmation only. The higher error from Stefan-Boltzmann is due primarily to the significant contribution to the error budget from the bolometric correction uncertainty. The error on this radius cannot be further reduced by a weighted average of the results from the two approaches, because their errors are highly correlated, both having contributions from the uncertainties in $T^{*}_{eff}$ and M$_V^{*}$.

\subsection{Notes on the Radii of Other PNNi}\label{notes-MV}
All quantities required to estimate the radii for our other PNNi, as we did for DeHt\,5, are gathered in Table~\ref{tbl-AQ}, where we also list our derived radii.

{\bf Abell\,31} - \cite{Kal83} estimates ``c" =0.0 $\pm$0.4. We note that the line of sight total extinction estimated by \cite{Sch98} is E($B$  $-$  $V$) = 0.064. In this case we  derive an average absorption per 100 pc for the four astrometric reference stars nearest in angular distance to the PNN
of A$_V^* = 0.05\pm0.04$. From \cite{Rau03} we calculate an intrinsic ($B$  $-$  $V$)$_{\rm o}$ = $-$0.36. This yields A$_V^* = 0.24 \pm 0.07$. The average of these methods yields A$_V^* = 0.10 \pm 0.07$  which we adopt. The distance modulus and V$_0$ provide an absolute magnitude M$_V = 6.31^{+0.29}_{-0.26}$, where we have included an LKH correction (-0.14 $\pm$ 0.03) and its uncertainty and the small uncertainty in A$_V^*$ in quadrature. 

\cite{Nap99} has estimated T$^{*}_{eff} = 84,700 K \pm$ 4,700 K.   From our synthetic photometry we calculate a B.C. = - 6.29 $\pm$ 0.2. As in Section~\ref{RAD}, above, we compare bolometric luminosities with the Sun and find $R_{*} = 0.041 \pm 0.017 R_{\sun}$. We obtain $H_V^{*}$ for $T^{*}_{eff}=84,700$ K, again from our synthetic photometry, which yields $R_{*}=0.039 \pm 0.006 R_{\sun}$.

{\bf NGC\,7293} -  has  a reliably low value for the extinction.
\cite{Kal83} finds a value of c = 0.04. Milne \& Aller (1975) derive
E($B$  $-$  $V$) = 0.01 from  H$\beta$ and radio. \cite{Boh82} using
IUE observations derives E(B-V)=0.012+/-0.03. \cite{Har07} quotes
E(B-V)=0.03 from $BVI$ photometry of the central star (assuming 
($B$  $-$  $V$)$_{\rm o}$ = $-$0.38), while Pottasch (1996) \nocite{Pot96} uses the same
technique and gets E($B$  $-$  $V$) = 0.0. The extinction maps of \cite{Bur82} and \cite{Sch98} would estimate a reddening of 
E($B$  $-$  $V$) $\sim$ 0.03 mag, consistent with the NaI measurements for the 
central star \citep{Mau04}. If we translate the values for $c$ to 
E($B$  $-$  $V$), and incorporate all of the other values we get a mean of 
$\langle$E($B$  $-$  $V$)$\rangle$ = 0.027 $\pm$ 0.022. Consistent with this value, we derive an average absorption per 100 pc from the three astrometric reference stars nearest in angular distance to the PNN. With a per-unit 100 pc $\langle$A$_V\rangle$=0.02$\pm$0.02 and the measured distance to the NGC\,7293 central star, d = 216$\pm 13$ pc, we obtain a total absorption for the PNN, A$_V^* = 0.04\pm0.03$. Again we adopt an average of the estimates,  A$_V^* = 0.09\pm0.04$, yielding M$_V = 6.74\pm 0.13$, where we have included an LKH correction (-0.03 $\pm$ 0.01) and its uncertainty and the small uncertainty in A$_V^*$ in quadrature. 

\cite{Nap99} has estimated T$^{*}_{eff} = 103,600 K \pm$ 5,500 K, which yields B.C. = -6.77.   Comparing bolometric luminosities with the Sun we obtain $R_{*} = 0.028 \pm 0.007 R_{\sun}$. We obtain $H_V^{*}$ for $T^{*}_{eff}=103,600$ K from our synthetic photometry, which yields $R_{*}=0.028 \pm 0.003 R_{\sun}$.  

{\bf NGC\,6853} -  To estimate A$_V$ \cite{Bar84}
found c = 0.17 and \cite{Kal76}
find c = 0.02. \cite{Boh82} estimate E(B-V)=0.06+/-0.03 from IUE
observations, while Pottasch (1996) gets E($B$  $-$  $V$) = 0.10 and \cite{Har07} find E($B$  $-$  $V$) = 0.07 from central star photometry.  \cite{Cia99} obtain c=0.11. The mean 
with error is then $\langle$E($B$  $-$  $V$)$\rangle$ = 0.08 $\pm$ 0.06, or $\langle$A$_V\rangle$=0.26 $\pm$ 0.17. Given this uncertain $\langle$A$_V\rangle$, we  derive an average absorption per 100 pc for the three stars nearest the central target, ref-4, -5, and -8. A finder chart can be found in \cite{Ben03}. With a per-unit 100 pc $\langle$A$_V\rangle$=0.07 $\pm$ 0.03 from these three stars and the measured distance to the NGC\,6853 central star, d = 405$^{+28}_{-25}$ pc, we obtain a total absorption for the PNN, A$_V^* = 0.28\pm0.06$. We adopt an average A$_V^* = 0.30\pm0.06$ and obtain M$_V = 5.62 \pm 0.16$, where we have included an LKH correction (-0.03 $\pm$ 0.01) and its uncertainty and the 0.06 magnitude uncertainty in A$_V^*$ in quadrature. 

 \cite{Nap99} has estimated T$^{*}_{eff} = 108,600 K \pm$ 6800 K for which we obtain B.C. = -6.91 $\pm$ 0.2, where the error is dominated by the uncertainty in the temperature and the behavior of the B.C. at these high temperatures. Comparing bolometric luminosities with the Sun
we find $R_{*} = 0.046 \pm 0.012 R_{\sun}$. Comparing $H_V$ with Vega we obtain $H_V^{*}$ for $T^{*}_{eff}=108,600$ K listed in Table~\ref{tbl-AQ}, which then yields $R_{*}=0.045 \pm 0.004 R_{\sun}$. Given that the two approaches yield $R_{*}$ values that agree within their errors, we adopt $R_{*}=0.045 \pm 0.004 R_{\sun}$.

\subsection{Astrophysical Consequences}\label{AC}

\subsubsection{Radii: PNNi vs.\ WDs}

Our four parallaxes, along with measured temperatures and apparent luminosities have resulted in four newly estimated radii for PNNi that, according to theory, should eventually descend to a WD cooling track. Previous investigations have yielded precise temperatures and radii (and masses) for five WD in visual, spectroscopic, and eclipsing binaries.  These results (Sirius B, Holberg et al. 1998\nocite{Hol98}; Procyon B, Girard et al. 2000\nocite{Gir00}; 40 Eri B, Shipman et al. 1997\nocite{Shi97}) are collected in \cite{Pro02}. To these we add Feige 24 \citep{Ben00a}, which has $\sigma_{\pi}/\pi = 2.7$\%, hence, a reasonably well-determined radius, two PNNi from \cite{Har07}, Sh 2-216 and HDW 4, both with $\sigma_{\pi}/\pi \le 8$\%, and the eclipsing Hyades WD binary, V471 Tau \citep{OBr01} . We will now compare PNNi and WD, using the quantities gathered in Table~\ref{tbl-comp}. We can also test the accuracy of the parallax and the many corrections leading to the bolometric magnitudes required by Equation 6, relating temperature, bolometric magnitude and radius. This aggregate of data probes the transition from PNNi to WD.

Let us for the moment assume that all PNNi have the same mass and radius. In such a Universe the only PNNi and/or WD variable is age, hence, temperature, and there should be a relationship between absolute bolometric magnitude and temperature, through Stefan-Boltzmann, L=4$\pi R^2\sigma T^4$. Figure~\ref{fig-6} (basically an HR diagram) indicates that temperature and absolute bolometric magnitude are correlated, in that hotter PNNi and WD have brighter absolute magnitudes. This is not surprising, given that the measured or estimated mass of most of these objects lie in the range $0.44 < {\cal M}_{\sun}< 0.60$. For the WD, bolometric magnitudes are derived using B.C. from Flower (1996) and \cite{Ber95}. The most discrepant object is Sirius B, which has the greatest  mass, ${\cal M}^*= 1.02{\cal M}_{\sun}$, hence, the smallest WD radius. 
If Equation 6 perfectly describes both PNNi and WD, then the residuals, $\Delta$M$_{bol}$, in Figure~\ref{fig-6} to our simple linear relationship between M$_{bol}$ and log T should correlate with the log of the radius. Figure~\ref{fig-7}, wherein  $\Delta$M$_{bol}$ is plotted against  log radius, shows such a correlation.  The residuals in Figure~\ref{fig-7} have an RMS dispersion of 0.12 magnitudes, indicating the accuracy of the LKH and extinction corrections. The two most discrepant objects are Feige\,24 and Sirius\,B with residuals of order 0.2 magnitudes.
\subsubsection{PNNi Masses and Gravities}
Various investigators have modeled the evolution of a star as it passes through the red giant phase, ejects significant mass, and becomes a white dwarf. The PN phase lies between the giant and WD stages in stellar evolution. 
To estimate PNNi mass we compare in Figure~\ref{fig-Mvall} the positions of our PNNi in an H-R diagram (M$_V$ - log T$_{eff}$) with predicted evolutionary tracks of post-AGB stars from Sch{\"o}nberner \& Bl{\"o}cker (1996). Also plotted are tracks of lower mass stars from \cite{Dri99}. Absolute V-band magnitudes, M$_V$, for tracks were calculated for solar metallicity, using SEDs from Rauch (2003). PNNi temperatures are from \cite{Nap99}. We find that the four PNNi clump about a mass ${\cal M} = 0.57 {\cal M}_{\sun}$, in agreement with the peak of the WD mass distribution, 0.60  ${\cal M}_{\sun}$, found by \nocite{Lie05} Liebert, Bergeron \& Holberg (2005). Interpolated individual masses are found in Table~\ref{tbl-AQ}.
We also estimate masses for PNNi HDW\,4 and Sh2-216. These are listed in Table~\ref{tbl-comp}. We note that the mass of 40 Eri B from these tracks is $\sim0.9{\cal M}_{\sun}$, differing substantially from that listed in \cite{Pro98}. 

A mass and radius uniquely determine a surface gravity, $g$, through
\beq
g = MG /R^2
\eeq
where G is the gravitational constant. Our DeHt\,5 radius, $R_{*}=0.025 \pm 0.002 R_{\sun}$ and the  mass inferred from the evolutionary tracks in Figure~\ref{fig-Rall} (where we plot radius vs. T$_{eff}$), ${\cal M} = 0.57 \pm 0.02{\cal M}_{\sun}$, yield log $g$ = 7.41$\pm$0.08.  The uncertainty in our log $g$ is primarily due to the radius uncertainty.
For Abell\,31 our mass estimate,  ${\cal M} = 0.53 \pm 0.03{\cal M}_{\sun}$ with our radius determines a gravity, log $g$ = 6.96$\pm$0.14. 
 For NGC\,7293 using the radius from the $H_V^{*}$ approach and a mass estimated from Figure~\ref{fig-Mvall},  ${\cal M} = 0.60 \pm 0.02{\cal M}_{\sun}$  determines a gravity, log $g$ = 7.34$\pm$0.06. 
For NGC\,6853 our mass estimate,  ${\cal M} = 0.57 \pm 0.01{\cal M}_{\sun}$ with our radius determines a gravity, log $g$ = 6.89$\pm$0.08, in agreement with the \cite{Nap99} line profile fitting value, log $g$ = 6.7$\pm$0.2.  Calculated values of log $g$ for our four PNNi are listed in Table~\ref{tbl-AQ} and compared with the \cite{Nap99} line profile fitting values.
Note that Figure~\ref{fig-Rall} shows 40 Eri B to have a mass consistent with past estimates. In Figure~\ref{fig-8} we compare our WD sample with the PNNi listed in Table~\ref{tbl-comp} on a mass-radius diagram. Most of the PNNi and the hottest WD, Feige 24, lie  above the  zero-temperature mass-radius relationship from \cite{Ham61}. These radii confirm  that PNNi are larger than cooler WD.

\subsubsection{Limits on Stellar Companions} \label{BINLIM2}
With the parallaxes from Table~\ref{tbl-SUM}, absolute magnitudes and mass estimates from Table~\ref{tbl-AQ}, and the separation and $\Delta m$ detection limits from Section~\ref{BINLIM}, we can now estimate the spectral type, separation in AU and periods for companions at the limit of detectability. Companions with spectral types later than listed in Table~\ref{tbl-BINLIM} would not be detected by the FGS. For example a companion two magnitudes fainter than the DeHt\,5 PNN (an M1V star) is at the limit of detectability for a separation of 15 mas, which for the parallax of DeHt\,5 equates to 5.2 AU.  From $P^2(M_1+M_2)=a^3$ we derive a period P=11y. For another application of Table~\ref{tbl-BINLIM} we consider NGC\,7293. \cite{Su07} find evidence for IR-emitting dust 35--150 AU from the PNN of NGC\,7293. A binary companion could provide an engine to sculpt a dust disk. Given that 35 AU at the distance of NGC\,7293 is 0\farcs16 and that we would detect a companion with  $\Delta m \le 3.5$, that companion would have to have a spectral type later than M1\,V. If much asymmetrical PN structure is due to binarity \citep{Sok06,DeM09}, then we would have the highest probability of detecting the companion to the PNN of NGC\,6853, the most asymmetric of the PN we observed. Our null detection suggests that such a companion is likely to have P$<$7y and ${\cal M} < 0.7{\cal M}_{\sun}$.

Finally, in principle we can probe for very short period companions using the photometry in Figure 2.
Companions with small separations could produce a single-peaked
orbital light curve through heating of the companion star by the PNN
e.g., the reflection effect. As the companion star orbits the PNN, its heated face is alternately more or less visible,
increasing and decreasing the observed flux from the PNN
once per orbit.  For example, the Feige 24 system (WD+M2\,V) has a period of 
4.23$^d$ and from FGS photometry (Benedict et al. 2000) evidences a photometric variation of 25 mmag.  \cite{Kaw08} estimate an inclination $i = 77$\arcdeg. Assuming a similar inclination, we find from binary light curve modeling
(cf. Harrison et al. 2009\nocite{Har09})  that even the
shortest period companions would
produce less V band variation from
reflection effects than we see in Figure 2. PNNi are too bright, washing out any variations due to reflection from companions.


\subsubsection{Comparison with other Distance Estimates} \label{Dist}
Napiwotzki (2001) has compiled PNe distances from a number of methods. Figure~\ref{fig-dist} plots our distances from Table~\ref{tbl-AQ} against three other methods; distances obtained via non-LTE PNNi atmosphere analysis (Napiwotzki 2001), from an H$\beta$-diameter relation devised by Shklovski (1956)\nocite{Shk56}, and from an interstellar Na D line analysis \citep{Nap95}. Shklovskii distances are only available for only three of our targets, and the Na D method has been applied only to two of our targets. The recent recalibration of the Shklovski distances by \cite{Sta08} left those three distances basically unchanged. The dashed line represents perfect agreement.  The Shklovski approach has a tendency to underestimate the distances while the spectroscopic distances are a bit on the high side. We now provide a more detailed assessment of distances obtained via non-LTE PNNi atmosphere analysis. 

\subsubsection{Reassessing the Spectroscopic Distance Scale for PNNi} \label{dcomp}
Napiwotzki (1999) determined PNNi fundamental stellar parameters, temperature and surface gravity, from a fit of the hydrogen Balmer lines with profiles computed from NLTE model atmospheres. This technique is well established and tested for the analysis of hot white dwarfs (e.g.\ Finley et al.\ 1997\nocite{Fin97}) using LTE atmospheres. However, when Napiwotzki (1999) applied this method to the even hotter central stars of old PNe, it became clear that for many stars no consistent fit of all Balmer lines could be achieved. A strong temperature trend was present, with the fit of higher members of the Balmer series yielding higher temperatures. This became known as the Balmer line problem (Napiwotzki \& Rauch 1994\nocite{Nap94}). 

\cite{Nap93, Nap99} presented arguments that the temperature derived from the highest Balmer lines H$\delta$ and H$\epsilon$ are close to the real temperatures of the PNNi. However, a physical explanation of the Balmer line problem remained elusive for some time. Models used for the Napiwotzki (1999) investigation were calculated in full NLTE but included only the two most abundant elements hydrogen and helium. Tests carried out prior to the start of this project appeared to show that the impact of line blanketing of heavier elements on the temperature structure of the atmospheres had only minor impact on the hydrogen line profiles (see discussion in Werner 1996). However, Werner (1996) could show that strong cooling by the resonance lines of carbon, nitrogen, and oxygen can reproduce the observed Balmer lines in the hot sdO BD+28$^\circ$4211 and the PNN of Sh\,2--216, if detailed treatment of the Stark broadening of these lines is included in the calculations. This treatment is a very computer time intensive. For this reason it was not included in previous calculations. The Werner (1996) \nocite{Wer96} results provided an astrophysical explanation and essentially validated the recipe of adopting the temperature fitted for H$\delta$ and H$\epsilon$ put forward in Napiwotzki (1993). This method was then adapted by Napiwotzki (1999). 

Fitted surface gravities appeared to be unaffected by the Balmer line problem. All Balmer lines could be fitted with a single value of $g$ using the Napiwotzki (1999) H and He models. Also, the Werner (1996) calculations did not indicate offsets in gravity. Napiwotzki (2001) compared his spectroscopic distance estimates with the results of other distance estimates including the best parallax measurements available at that time \citep{Har97}. The comparison showed spectroscopic distances larger than trigonometric distances by 55\% -- at face value. However, as pointed out by Napiwotzki (2001), one has to take into account that the often significant relative errors of trigonometric parallaxes introduce sample biases. Lutz--Kelker biases are one way to estimate the size of the effect. Napiwotzki (2001) performed a Monte--Carlo simulation trying to model the properties of the  \cite{Har97} sample as closely as possible given the -- by necessity -- not well defined selection criteria. The result was an estimated bias of the trigonometric distances, now too small by $32\pm 0.25$\%. The conclusion at that time was that both distance scales are marginally consistent, but large uncertainties remained. 

Improved accuracy of recent trigonometric parallax measurements have changed the situation dramatically. \cite{Har07} achieved accuracies ranging from 0.3\,mas to 0.6\,mas. The measurements presented in our investigation yield even better accuracies with $\sigma_\pi$ being 0.21\,mas and below. \cite{Har07} estimates a bias of 5\% for their sample. A straightforward reading from Table~A.2 in Napiwotzki (2001) gives an estimate of 7\% for a sample with 0.2\,mas accuracy. Both estimates confirm that remaining systematic errors are now much smaller, in line with the small Lutz--Kelker biases given in our Table~\ref{tbl-AQ}. Thus it is a good time to re-assess the spectroscopic distance scale.

Spectroscopic distances of the four program stars are 38\% larger than the trigonometric distances (unweighted average). This offset is smaller than that found by Napiwotzki (2001), but due to the smaller errors and biases it is now highly significant. This translates into an average $\log g$ offset $\overline{\log g_{\mathrm{spec}}-\log g_\pi} = -0.41$. In Table~\ref{tbl-AQ} we find DeHt\,5 anomalous, the difference between gravity derived from our radius and from the analysis of stellar atmospheres  larger than for the other three PNNi.  DeHt\,5 is of special interest and will be discussed below. Excluding DeHt\,5 from the average we compute an average distance offset of 40\% (spectroscopic
   distances recomputed using the improved photometric observations,
   reddening estimates and synthetic photometry) 
and $\log g$ offset of 0.30, not much larger than typical gravity errors given in Napiwotzki (1999) which are 0.2--0.3 dex.

\subsubsection{The case of DeHt\,5}

The temperature and gravity derived by Napiwotzki (1999) place the central star of DeHt\,5 in a region of the temperature gravity diagram inconsistent with a post-AGB origin. The  parameters of this central star were better matched to those of a star which lost its envelope at the end of the first red giant branch (RGB) and is now evolving into a low mass He-core white dwarf. \cite{Bar01} performed an analysis of optical and UV spectra using model atmospheres including the effect of model line blanketing. The derived temperature is 57,400\,K -- lower than the Napiwotzki (1999) result -- partly due to the inclusion of metal line blanketing and partly due to a different fit algorithm and philosophy. The gravity $\log g=7.0$ given by \cite{Bar01}  is essentially an upper limit, because it was the lowest $\log g$ available in the model atmosphere grid. The gravity resulting from a fit with pure hydrogen models is in good agreement with the Napiwotzki (1999) result: $\log g = 6.75$ vs. 6.65. 

None of the results can be reconciled with the trigonometric results, which translates into $\log g = 7.4$. \cite{Bar01} also determined metal abundances from the analysis of an HST--STIS UV spectrum. The resulting abundances are higher than those of the well-known hot ``template'' white dwarf G\,191--B2B, which appears to have typical abundances for that parameter range (Barstow et al. \ 2003). One could be tempted to speculate that unusual metal abundances could explain the unusual large differences between trigonometric and spectral analysis results. More detailed abundance analyses of more PNNi would be needed to decide this question. In any case one conclusion is that even the metal blanketed atmospheres of \cite{Bar01} are not capable of producing results consistent with the trigonometric parallaxes.

The stellar parameters implied by the trigonometric parallax ($R=0.025R_\odot$ or $\log g=7.4$) places DeHt\,5 at a location expected for a run-of-the-mill pre-white dwarf. The implied mass  $M=0.57M_\odot$ sits spot on the main peak of the white dwarf mass distribution ($M=0.572$; Liebert et al. \ 2005). The implication is that DeHt\,5 is a rather normal C/O white dwarf resulting from post-AGB evolution. Somewhat problematic is the high implied post-AGB age of $>4\times 10^5$\,yr (read from the $0.605M_\odot$ track). The observational sample implies that PNe disperse into the intersteller medium after about $50-100\times 10^3$\,yr (e.g.\ Table~1 in Napiwotzki 2001). A post-AGB age as high as implied for DeHt\,5 would be highly unusual. \cite{Par06} speculate that DeHt\,5 is not a real PNe, but a chance association of an interstellar cloud with a hot white dwarf. A more detailed investigation of the nebula will clarify this issue.


\section{Summary}{\label{SUMRY}
{\it HST} FGS photometry indicates that none of these PNNi shows photometric variation larger than 5 mmag. From FGS interferometric fringe morphology we establish companion limits mid-K\,V to early M\,V. FGS interferometric fringe tracking astrometry yields  absolute trigonometric parallaxes for the PNNi of  DeHt\,5, Abell\,31, NGC\,7293, and NGC\,6853. Weighted averages with previous ground-based determinations \citep{Har07} provide parallaxes with errors at  or below 0.2 mas, or  $\langle \sigma_{\pi}/\pi\rangle = 5$\%. Our results confirm that statistical distances
methods of the Shklovski type underestimate the distances of old planetary nebulae. On the other hand, the improved accuracy of our trigonometric parallaxes now show that previous spectroscopic distances significantly overestimated the distances. We see a consistent trend in the spectroscopic distance scale overestimating the true distance by 40\% (corresponding to an understimate of $\log g$ by 0.3\,dex). Results from Napiwotzki (1999) and similar studies should be corrected accordingly.  We use these parallaxes and estimates of interstellar extinction from our own spectrophotometry and other investigations to derive PNNi absolute magnitudes. With these we derive radii, either comparing with the Sun through bolometric corrections, or with Vega via the V-band average flux, $H_V$. These four PNNi along with two others with well-determined distances and five WD satisfy theoretical linear correlations between absolute bolometric magnitude, log temperature, and log radius.  Estimating from post-AGB evolutionary models, we find PNNi masses that agree with those typically found for white dwarf stars. The PNNi and the hottest WD clearly fall above a WD mass-radius relationship established by nearby, cool WD.

\acknowledgments

Support for this work was provided by NASA through grants NAG5-1603, GO-10432, and GO-10611 from the Space Telescope 
Science Institute, which is operated
by AURA, Inc., under
NASA contract NAS5-26555. These results are based partially on observations obtained with the
Apache Point Observatory 3.5 m telescope, which is owned and operated by
the Astrophysical Research Consortium. Washington/DDO photometry was secured at Las Campanas Observatory (Carnegie Institute of Washington)  and Fan Mountain Observatory (University of Virginia). This publication makes use of data products from the Two Micron All Sky Survey,
which is a joint project of the University of Massachusetts and the Infrared Processing
and Analysis Center/California Institute of Technology, funded by NASA and the NSF. 
This research has made use of the SIMBAD database and {\it Aladin}, both developed at CDS, Strasbourg, France; the NASA/IPAC Extragalactic Database (NED) which is operated by JPL, California Institute of Technology, under contract with the NASA;  and NASA's Astrophysics Data System Abstract Service. We thank an anonymous referee for a careful reading and suggestions that improved the presentation.


\bibliography{my.bib}

\clearpage

\begin{deluxetable}{lllrrr}
\tablewidth{7in}
\tablecaption{PNNi Positions and Aliases\label{tbl-PN}}
\tablehead{\colhead{PN}&
\colhead{RA~~~~(2000)}&\colhead{Dec}& \multicolumn{3}{c}{Aliases} 
}
\startdata
NGC\,6853&19 59 36.34&+22 43 16.1&Dumbbell&M 27&WD 1957+225\\
NGC 7293&22 29 38.55&$-$20 50 13.6&Helix&PN G036.1$-$57.1&WD 2226$-$210\\
DeHt\,5&22 19 33.71&+70 56 03.3& &PN G111.0+11.6&WD 2218+706\\
Abell 31&08 54 13.16&+08 53 53.0&PN A66 31&PK 219+31&PN CSI+09$-$08515\\
\enddata
\end{deluxetable}

\begin{deluxetable}{llllccccccccccc}
\tablewidth{7in}
\tablecaption{DeHt\,5 Log of Observations and Reference Star Availability (x = Observed)\label{tbl-LOO}}
\tablehead{\colhead{Set}&
\colhead{mJD}&\colhead{P$_{\alpha}$\tablenotemark{a}}&\colhead{P$_{\delta}$\tablenotemark{b}} &\colhead{DeHt5}&\colhead{2\tablenotemark{c}}&\colhead{3}&\colhead{4}&\colhead{5}&\colhead{6}&\colhead{7}&\colhead{8}&\colhead{9}&\colhead{10}&\colhead{11}
}
\startdata
1&53390.0556&$-$0.5020&$-$0.8415&x&x&x&x&x&&x&x&x&x&x\\
2&53392.0128&$-$0.4752&$-$0.8578&x&x&x&x&x&&x&x&x&x&x\\
3&53587.7651&0.3155&0.9660&x&x&x&x&x&&x&x&x&x&x\\
4&53587.8598&0.3147&0.9663&x&x&x&x&x&&x&x&x&x&x\\
5&53623.0390&$-$0.2369&0.9536&x&x&x&&x&x&&&&&\\
6&53671.1814&$-$0.8217&0.4197&x&x&x&x&&x&&&&&\\
7&53764.2406&$-$0.3764&$-$0.9099&x&x&x&x&x&&x&x&x&x&\\
8&53772.1216&$-$0.2563&$-$0.9501&x&x&x&&x&&x&x&x&x&x\\
9&53783.0153&$-$0.0818&$-$0.9759&x&x&&&x&x&&&&&\\
10&53956.0037&0.2697&0.9764&x&x&x&x&x&x&&&&&\\
11&53957.1156&0.2521&0.9805&x&x&x&x&x&x&&&&&\\
12&54780.6123&$-$0.8997&0.1901&x&x&x&x&&x&&&&&\\
13&54780.6788&$-$0.9000&0.1890&x&x&x&x&&x&&&&&\\
14&54780.7454&$-$0.9003&0.1879&x&x&x&x&&x&&&&&\\
15&54782.5436&$-$0.9067&0.1574&x&x&x&x&&x&&&&&\\
\enddata
\tablenotetext{a}{Parallax factor in Right Ascension }
\tablenotetext{b}{Parallax factor in Declination}
\tablenotetext{c}{Reference star number}
\end{deluxetable}

\begin{deluxetable}{llllllll}
\tablewidth{0in}
\tablecaption{DeHt\,5 Astrometric Reference Star Photometry, Spectral Classifications, and
Spectrophotometric Parallaxes \label{tbl-SPP}}
\tablehead{\colhead{ID}& \colhead{V}&
\colhead{B-V} & \colhead{V-K} & \colhead{SpT} & \colhead{M$_V$} & \colhead{A$_V$} &
\colhead{$\pi_{abs}$}(mas) } 
\startdata
Ref-2&15.56&0.82&2.29&F5\,V&3.34&1.33&0.66$\pm$0.15\\
Ref-3&14.90&0.82&2.2&F4\,V&3.11&1.33&0.77$\pm$0.19\\
Ref-4&14.31&0.96&2.54&G2\,V&4.56&1.25&1.98$\pm$0.46\\
Ref-5&11.92&0.48&1.24&F4\,V&3.11&0.25&1.84$\pm$0.45\\
Ref-6&13.55&0.6&1.63&F4\,V&3.11&0.72&1.07$\pm$0.26\\
Ref-7&15.02&1.06&2.76&G2\,V&4.56&1.49&1.60$\pm$0.37\\
Ref-8&14.90&0.83&2.72&F2\,V&2.84&1.32&0.85$\pm$0.17\\
Ref-9&13.50&1.42&3.75&K1\,III&0.6&1.34&0.49$\pm$0.12\\
Ref-10&14.94&0.92&2.52&F7\,V&3.72&1.27&1.05$\pm$0.24\\
Ref-11&14.86&1.48&3.88&K2\,III&0.5&1.34&0.25$\pm$0.06\\
\enddata
\end{deluxetable}

\begin{deluxetable}{lccccc}
\tablewidth{0in}
\tablecaption{DeHt\,5  and Reference Star Astrometric Data    \label{tbl-POS}}
\tablehead{\colhead{ID}&
\colhead{$\xi$ \tablenotemark{a}} &
\colhead{$\eta$ \tablenotemark{a}} &
\colhead{$\mu_x$ \tablenotemark{b}} &
\colhead{$\mu_y$ \tablenotemark{b}}&\colhead{$\pi_{abs}$\tablenotemark{c}} }
\startdata
DeHt\,5\tablenotemark{d}&$-$4.5797$\pm$0.0001&$-$2.5814$\pm$0.0001&$-$11.80$\pm$0.10&$-$18.49$\pm$0.08&2.86$\pm$0.16\\
Ref$-$2&$-$3.5388$\pm$0.0002&\phs52.3476$\pm$0.0001&$-$3.19$\pm$0.11&$-$5.59$\pm$0.10&0.70$\pm$0.05\\
Ref$-$3&$-$2.0182$\pm$0.0002&$-$156.6239$\pm$0.0002&$-$2.51$\pm$0.12&$-$5.21$\pm$0.13&0.90$\pm$0.03\\
Ref$-$4&$-$92.2348$\pm$0.0003&$-$92.3332$\pm$0.0002&\phs1.73$\pm$0.12&\phs1.77$\pm$0.16&1.30$\pm$0.04\\
Ref$-$5&\phs53.3040$\pm$0.0001&161.1489$\pm$0.0001&\phs0.20$\pm$0.24&$-$7.28$\pm$0.28&1.50$\pm$0.05\\
Ref$-$6&\phs117.4994$\pm$0.0002&$-$52.1638$\pm$0.0002&\phs2.15$\pm$0.10&$-$4.14$\pm$0.10&1.10$\pm$0.08\\
Ref$-$7&$-$116.1855$\pm$0.0004&$-$189.3931$\pm$0.0004&\phs6.41$\pm$0.55&\phs5.39$\pm$0.64&0.80$\pm$0.15\\
Ref$-$8&$-$132.3795$\pm$0.0003&$-$311.4981$\pm$0.0004&\phs0.43$\pm$0.50&\phs0.10$\pm$0.47&1.21$\pm$0.10\\
Ref$-$9&$-$47.7228$\pm$0.0003&$-$263.5350$\pm$0.0003&$-$3.29$\pm$0.50&$-$1.06$\pm$0.49&0.34$\pm$0.16\\
Ref$-$10&\phs73.5099$\pm$0.0004&\phs89.0934$\pm$0.0003&\phs0.30$\pm$0.41&\phs1.17$\pm$0.51&1.01$\pm$0.08\\
Ref$-$11&$-$8.2819$\pm$0.0006&\phs202.4660$\pm$0.0007&$-$1.16$\pm$0.96&\phs1.07$\pm$1.02&0.21$\pm$0.16\\

\enddata
\tablenotetext{a}{$\xi$ (RA) and $\eta$ (Dec) are relative positions in arcseconds.
}
\tablenotetext{b}{$\mu_x$ and $\mu_y$ are relative motions in mas
yr$^{-1}$, where $x$ and $y$ are aligned with RA and Dec. }
\tablenotetext{c}{Absolute parallax in mas}
\tablenotetext{d}{RA = 22$^h$19$^m$33.713$^s$ +70\arcdeg56'03.28", J2000, epoch = mJD
53764.24692}
\end{deluxetable}

\begin{center}
\begin{deluxetable}{rllll}
\tablecaption{Reference Frame Statistics and PNNi Parallax and Proper Motion\label{tbl-SUM}} 
\tablewidth{7in}
\tablecolumns{5} 
\tablehead{\colhead{Parameter} & \multicolumn{4}{c}{PNNi}}
\startdata
& DeHt 5&Abell\,31&NGC 7293&NGC\,6853\\
\HST Study Duration (y)&3.81&3.99&1.84&9.10\\
Observation Sets (\#)&15&15&11&12\\
Ref stars (\#)&10&6&3&7\\
Ref stars $\langle$V$\rangle$&14.29&13.56&13.11&14.37\\
Ref stars $\langle$B-V$\rangle$&0.94&0.80&0.69&1.28\\
$<\sigma_\xi>$ (mas)&0.3& 0.3 & 0.9& 0.3 \\
$<\sigma_\eta>$ (mas)& 0.3 & 0.2 & 0.9&0.3\\
{\it HST} $\pi_{abs}$ (mas)&2.86$\pm$0.16&1.51$\pm$0.26&4.67$\pm$0.33&2.22$\pm$0.19\\
{\it HST} Relative $\mu$ (mas y$^{-1}$)&21.93$\pm$0.12&10.49$\pm$0.13&38.99$\pm$0.24&8.70$\pm$0.11\\
in Position Angle (\arcdeg)&212.5$\pm$0.10&227.1$\pm$0.2&100.1$\pm$0.3&67.9$\pm$0.11\\
\\
{\it USNO} Ref stars (\#)&15&5&6&28\\
Ref stars (\#) in common&3&0&1&2\\
{\it USNO} $\pi_{abs}$ (mas)&3.34$\pm$0.56&1.76$\pm$0.33&4.56$\pm$0.49&3.17$\pm$0.32\\
{\it USNO} Relative $\mu$ (mas y$^{-1}$)&21.40$\pm$0.20&10.40$\pm$0.10&33.0$\pm$0.1&13.50$\pm$0.25\\
in Position Angle (\arcdeg)&214.6$\pm$0.5&226.5$\pm$0.6&86.7$\pm$0.3&60.8$\pm$1.0\\
\\
Weighted \HST+{\it USNO} $\pi_{abs}$&2.90$\pm$0.15&1.61$\pm$0.21&4.64$\pm$0.27&2.47$\pm$0.16\\
\enddata
\end{deluxetable}
\end{center}

\begin{tiny}
\begin{center}
\begin{deluxetable}{lllll}
\tablecaption{PNNi Astrophysical Quantities \label{tbl-AQ}}
\tablecolumns{5} 
\tablewidth{0in}
\tablehead{\colhead{Parameter} & \multicolumn{4}{c}{PNNi}}
\startdata
& DeHt 5&Abell\,31&NGC 7293&NGC\,6853\\
$ V $\tablenotemark{a}    &   15.47  &15.52 & 13.53 & 13.99 \\
\bv\tablenotemark{a} & -0.22 & -0.29 & -0.32 & -0.30 \\
d (pc)\tablenotemark{b} & 345$^{+19}_{-17}$& 621$^{+91}_{-70}$& 216$^{+14}_{-12}$& 405$^{+28}_{-25}$ \\
$A_V^*$ &$ 0.37 \pm 0.07 $ &$ 0.10 \pm 0.07 $&$ 0.09 \pm 0.04 $&$ 0.30 \pm 0.06 $ \\
m-M  & $7.69 \pm 0.12$ & $8.97 \pm 0.28$& $6.67 \pm 0.13$& $8.04 \pm 0.14$\\
LKH Bias & -0.02 $\pm$ 0.02  & -0.14 $\pm$ 0.03 & -0.03 $\pm$ 0.01 & -0.03 $\pm$ 0.01\\
$M_V$& 7.39 $\pm$ 0.14 & 6.31 $\pm$0.30 & 6.74 $\pm$0.13& 5.62 $\pm$ 0.16 \\
$T_{eff}^{*}$(K)\tablenotemark{c} & $76,500 \pm 5,800 $ & $84,700 \pm 4,700 $& $103,600 \pm 5,500 $& $108,600 \pm 6,800 $\\
B.C.& $-5.84 \pm 0.2$ & $-6.29 \pm 0.2$& $-6.77 \pm 0.2$& $-6.91 \pm 0.2$ \\
$M_{bol}^{*}$&  $+1.55 \pm 0.24$&  $+0.03 \pm 0.43$&  $-0.03 \pm 0.24$&  $-1.29 \pm 0.25$ \\
$H_V^{*}$\tablenotemark{d} &1.12$\times10^8$ &1.24$\times10^8$ &1.62$\times10^8$ &1.72$\times10^8$\\
$R_{*} (\sun) $& $ 0.025 \pm 0.002 $&  $ 0.039 \pm 0.006 $& $ 0.028 \pm 0.003 $& $ 0.045 \pm 0.004 $\\
${\cal M}_{*}(\sun)$\tablenotemark{e} & $ 0.57\pm 0.02$ & $ 0.53\pm 0.03$& $ 0.60\pm 0.02$& $ 0.57\pm 0.01$ \\
log $g$\tablenotemark{c}& 6.7 $\pm$ 0.2& 6.6 $\pm$ 0.3& 7.0 $\pm$ 0.2& 6.7 $\pm$ 0.2 \\
log $g$\tablenotemark{f}& 7.41 $\pm$ 0.08 & 6.99 $\pm$ 0.14& 7.34 $\pm$ 0.06& 6.89 $\pm$ 0.08 \\
\enddata
\tablenotetext{a}{from \cite{Har07}} \tablenotetext{b}{from weighted average of \HST and \cite{Har07}} \tablenotetext{c} {from \cite{Nap99}} \tablenotetext{d} {ergs cm$^{-2}$ s$^{-1}$ \AA$^{-1}$ str$^{-1}$}  \tablenotetext{e} {from Fig.~\ref{fig-Mvall}}\tablenotetext{f} {from $g = {\cal M}G /R^2$}
\end{deluxetable}
\end{center}
\end{tiny}

\begin{center}
\begin{deluxetable}{lccccllll}
\tablecaption{Comparing PNNi and WD \label{tbl-comp}} 
\tablewidth{0in}
\tablehead{\colhead{ID} &  \colhead{V$_0$} & \colhead{(m-M)$_0$} & \colhead{M$_V$\tablenotemark{f}}&  \colhead{BC} &  \colhead{M$_{bol}$}&  \colhead{log T (K)}&  \colhead{log R$^*(\sun)$}&  \colhead{{\cal M}$(\sun)$\tablenotemark{g}}}
\startdata
DeHt 5\tablenotemark{a}&15.15&7.69&7.39&-5.95&1.44$\pm$0.28&4.884$\pm$0.04&-1.61$\pm$0.03&0.57$\pm$0.02\\
Abell\,31\tablenotemark{a}&15.47&8.97&6.25&-6.3&-0.06 0.43&4.928 0.03&-1.40 0.06&0.53 0.03\\
NGC\,7293\tablenotemark{a}&13.50&6.67&6.74&-6.77&-0.03 0.24&5.015 0.03&-1.56 0.06&0.60 0.02\\
NGC\,6853\tablenotemark{a}&13.75&8.04&5.62&-7.09&-1.47 0.25&5.036 0.03&-1.35 0.04&0.57 0.01\\
Sh 2-216\tablenotemark{b}&12.38&5.54&6.82&-6.30&+0.52 0.23&4.920 0.03&-1.47 0.10&0.55 0.03\\
HDW 4\tablenotemark{b}&16.11&6.60&9.45&-4.47&4.98 0.28&4.674 0.03&-1.87 0.12&0.77 0.07\\
Feige 24\tablenotemark{c}&12.56&4.17&8.39&-4.82&3.57 0.13&4.751 0.02&-1.73 0.02&0.57 0.02\\
V471 Tau\tablenotemark{d}&13.72&3.39&10.33&-3.49&6.84 0.03&4.538 0.01&-1.97 0.01&0.84 0.05\\
Procyon B\tablenotemark{e}&10.82&-2.28&13.1&0&13.1 0.03&3.889 0.01&-1.91 0.01&0.55 0.02\\
Sirius B\tablenotemark{e}&8.44&-2.89&11.33&-2.3&9.03 0.1&4.394 0.01&-2.06 0.01\tablenotemark{h}&1.02 0.02\\
40 Eri B\tablenotemark{e}&9.5&-1.49&10.99&-1.5&9.49 0.1&4.223 0.01&-1.87 0.006&0.501 0.011\\
\enddata
\tablenotetext{a}{from this paper.}
\tablenotetext{b}{from  \cite{Har07}.}
\tablenotetext{c}{from  \cite{Ben00a}.}
\tablenotetext{d} {from  \cite{OBr01}. BC from \cite{Flo96} and \cite{Ber95}.}
\tablenotetext{e} {from compilation of \cite{Pro02}. BC from \cite{Flo96} and \cite{Ber95}.}
\tablenotetext{f} {includes LKH bias correction, negligible for the last four objects.}
\tablenotetext{g} {from this paper and \cite{Pro02}, except Feige 24 from \cite{Kaw08},  Procyon B and Sirius B from \cite{Sch06} and \cite{Bon09}.}
\tablenotetext{h} {from \cite{Bar05}.}
\end{deluxetable}
\end{center}

\begin{center}
\begin{deluxetable}{lccccllll}
\tablecaption{Companion Limits from FGS Fringe Scanning \label{tbl-BINLIM}}
\tablewidth{0in}
\tablehead{\colhead{ID} &  \colhead{M$_V$} & \colhead{$\pi_{abs}$ } & \colhead{${\cal M}_{*}(\sun)$}&  \colhead{$\Delta$V} &  \colhead{Comp 2 SpT}&  \colhead{Comp 2$ {\cal M}(\sun)$}&  \colhead{Sep [AU]}&  \colhead{P [y]}}
\startdata
DeHt\,5&7.39&2.9&0.57&1&K8V&0.59&3.4&6\\
&&&&2&M1V&0.53&5.2&11\\
&&&&3&M2V&0.4&17.2&73\\
\\
Abell\,31&6.23&1.61&0.53&1&K4V&0.7&6.2&14\\
&&&&2&K8V&0.59&9.3&27\\
&&&&3&M1V&0.53&31.1&168\\
\\
NGC\,7293&6.77&4.66&0.6&1&K6V&0.64&2.1&3\\
&&&&2&M0V&0.51&3.2&5\\
&&&&3&M1V&0.53&10.7&33\\
\\
NGC\,6853&5.62&2.47&0.57&1&K3V&0.72&4.0&7\\
&&&&2&K6V&0.64&6.1&14\\
&&&&3&M0V&0.51&20.2&88\\
\enddata
\end{deluxetable}
\end{center}

%
%

\begin{figure}
\epsscale{1.00}
\plotone{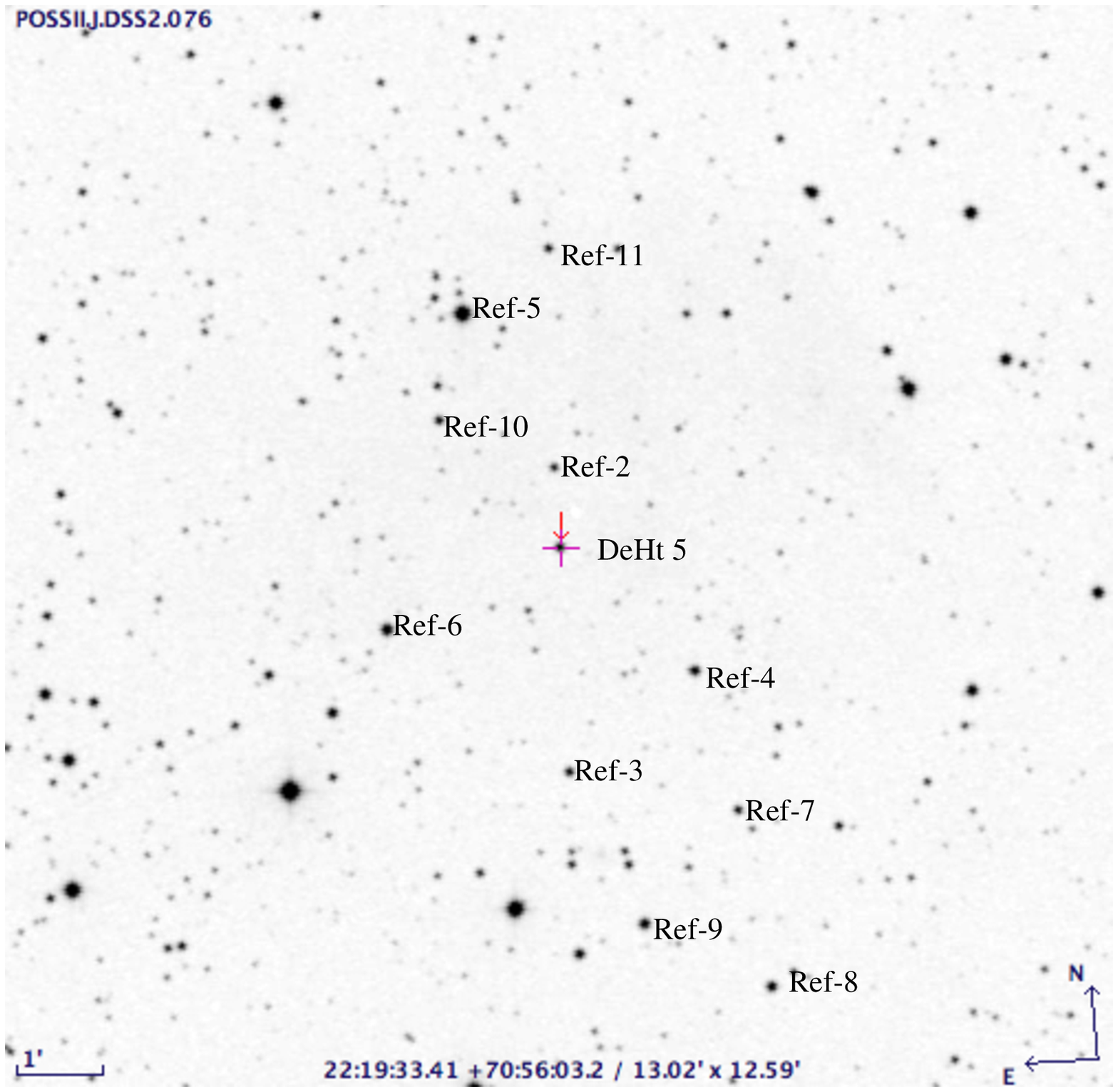}
\caption{DeHt5 central star and astrometric reference stars. Labels are immediately to the right of each star.}
\label{fig-1}
\end{figure}

\clearpage

\begin{figure}
\epsscale{0.85}
\plotone{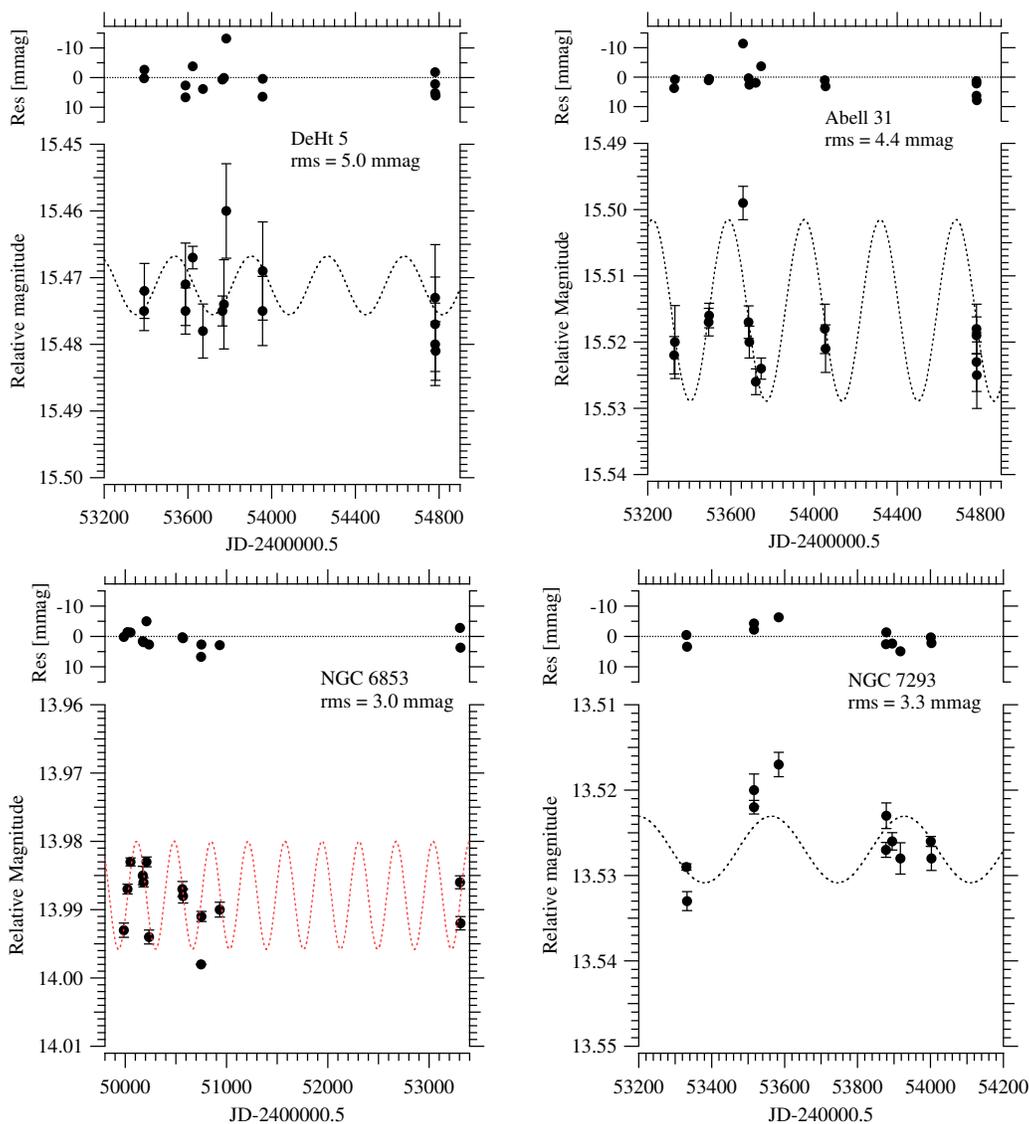}
\caption{Photometric variations of the four PNNi observed with the FGS. In each panel the bottom section shows a fit to the original variation ascribed to roll-induced one year period modulation of the stars used to generate each flat field. The top panel shows the residuals, each labeled with the final (presumed intrinsic) rms variation in the PNNi.} 
\label{FigLC}
\end{figure}
\clearpage

\begin{figure}
\epsscale{1.00}
\plotone{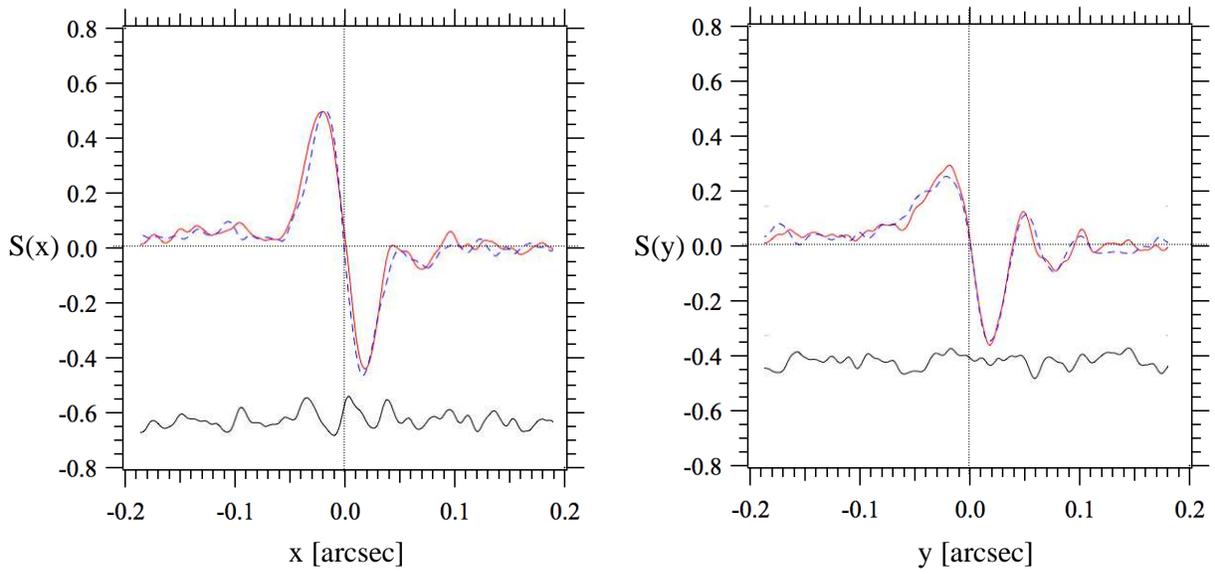}
\caption{Fringes along X and Y axes of DeHt\,5 (solid) and Abell\,31 (dashed) compared.
Residuals (bottom traces in each panel) indicate only the amplitude of noise expected for these
relatively faint targets.} 
\label{Fig-SC}
\end{figure}
\clearpage

\clearpage
\begin{figure}
\epsscale{0.75}
\plotone{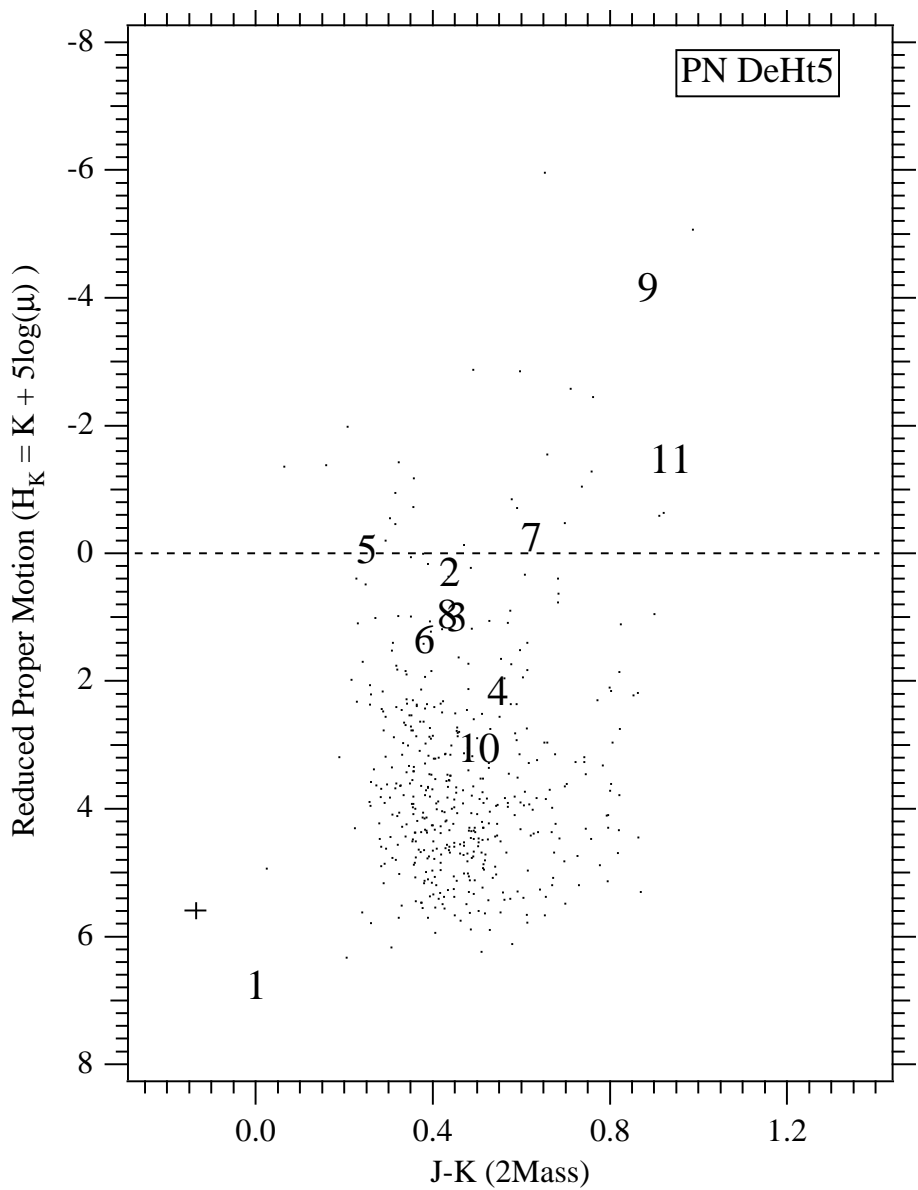}
\caption{Reduced proper motion diagram for 458 stars in a 1\arcdeg ~ field centered on DeHt 5. Star identifications are shown for DeHt 5 (=1) and our astrometric reference stars. H$_K$ for these stars is calculated using our final proper motions (Table~\ref{tbl-POS}). For a given spectral type giants and sub-giants have more negative H$_K$ values and are redder than dwarfs in J-K. Reference stars ref-9 and ref-11 are confirmed to be giant stars. The cross in the lower left corner indicates representative errors along each axis.}
\label{fig-RPM}
\end{figure}
\clearpage

\begin{figure}
\epsscale{1.00}
\plotone{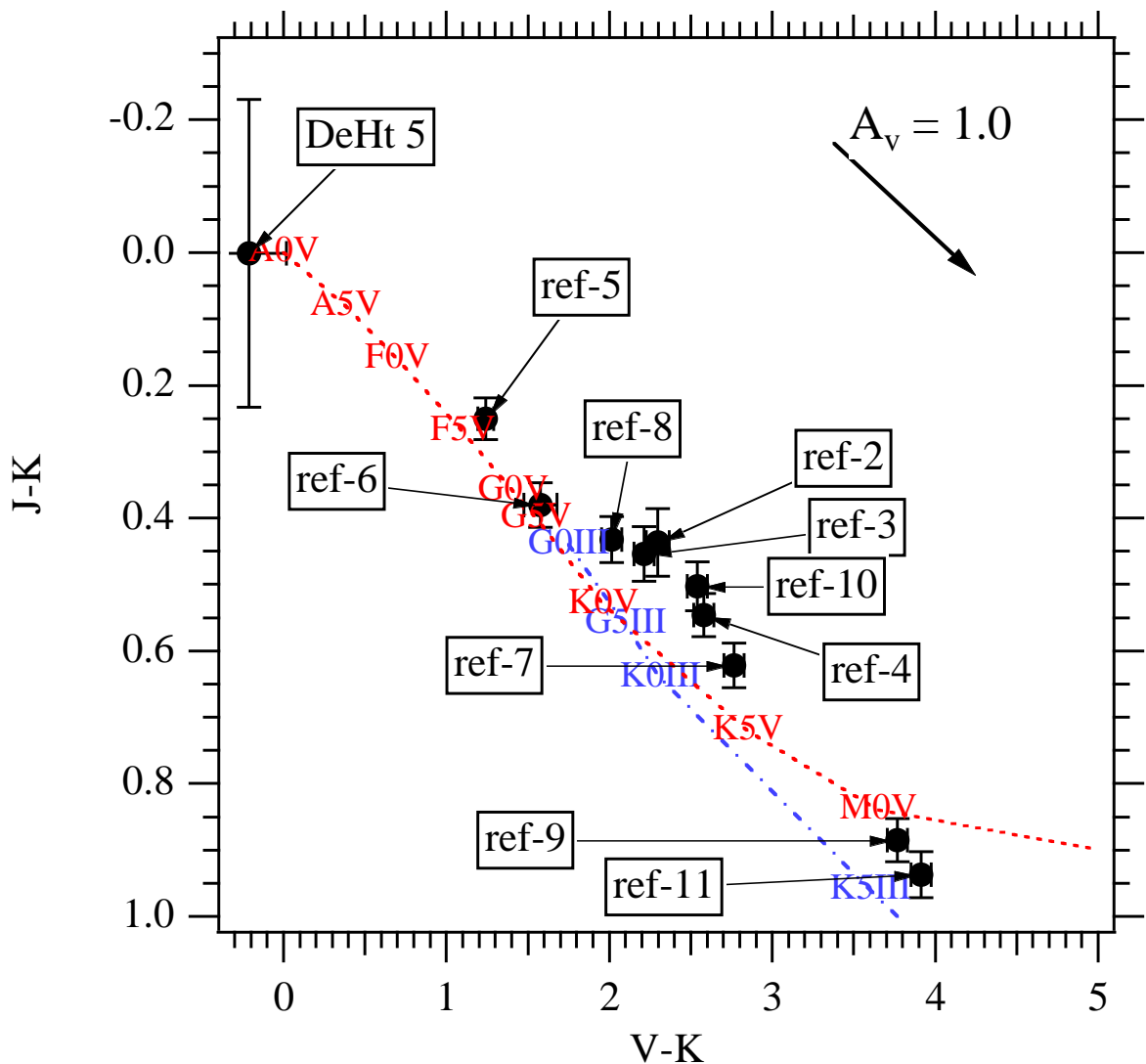}
\caption{ J-K vs V-K color-color diagram. The dashed line is the locus of  dwarf (luminosity class V) stars of various spectral types; the dot-dashed line is for giants (luminosity class III). The reddening vector indicates A$_V$=1.0 for the plotted color system.}
\label{fig-CCD}
\end{figure}

\begin{figure}
\epsscale{0.65}
\plotone{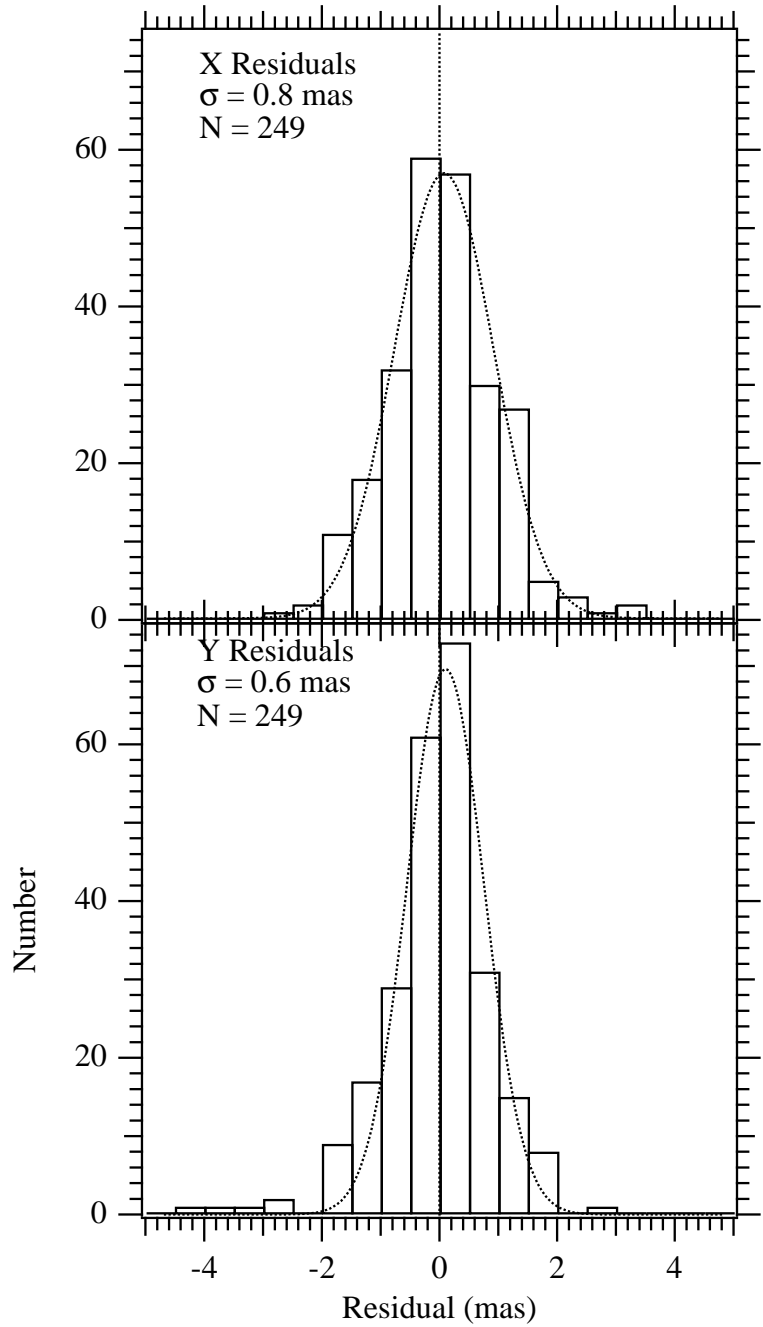}
\caption{ Histograms of x and y residuals obtained from modeling DeHt 5 and the astrometric reference stars with equations 4 and 5. Distributions are fit
with Gaussians whose $\sigma$'s are noted in the plots.} 
\label{fig-4}
\end{figure}
\clearpage

\begin{figure}
\epsscale{0.75}
\plotone{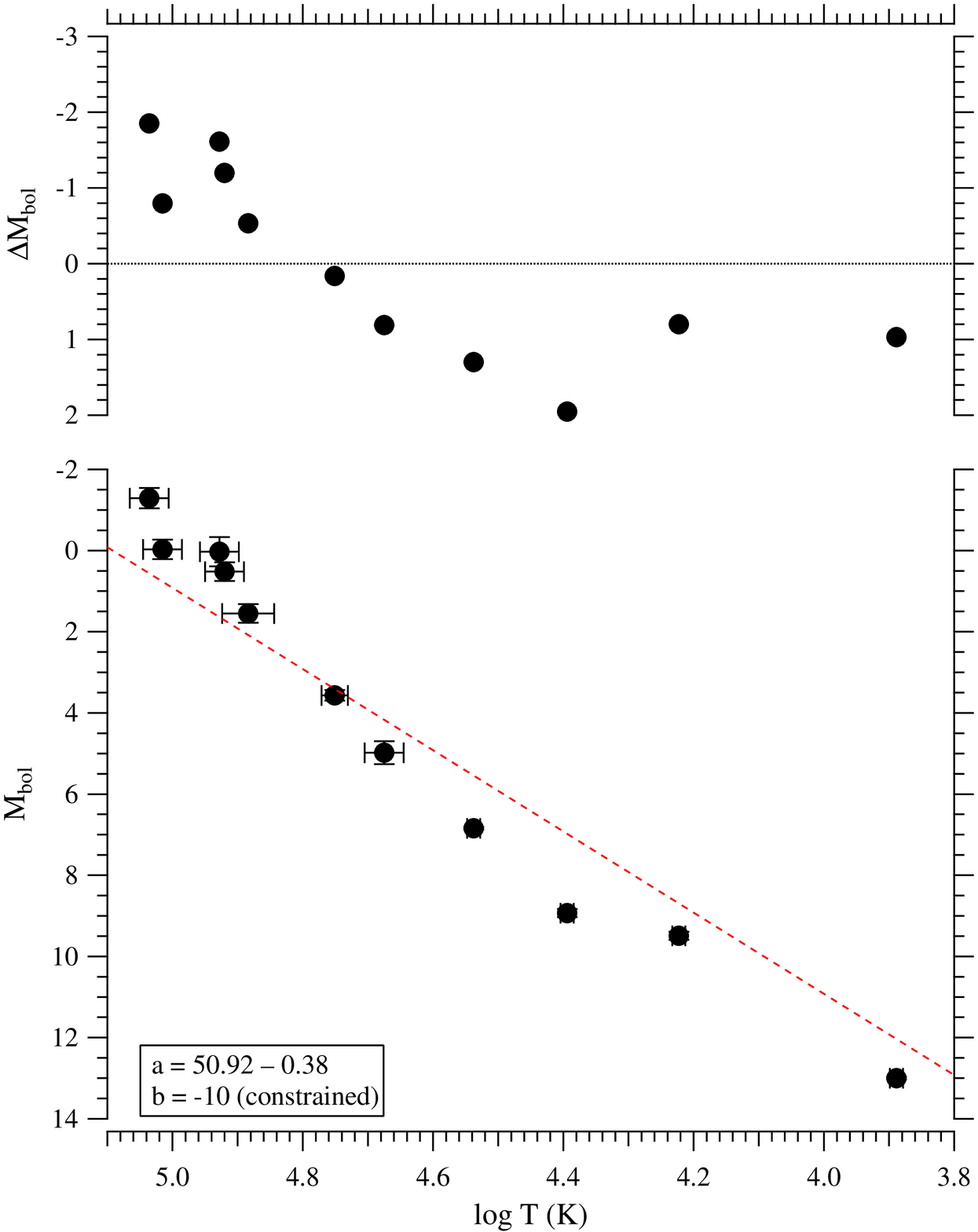}
\caption{Absolute bolometric magnitude, M$_{bol}$ plotted against log temperature. The slope is constrained to the theoretical value. M$_{bol}$ for the PNNi comes from our parallax, the apparent magnitude, the bolometric correction, and the interstellar absorption, A$_V$ (Table~\ref{tbl-AQ}) and \cite{Har07}. The WD values are from \cite{Ben00a} for Feige 24, from \cite{OBr01} for V471 Tau, and calculated from Provencal (2002), with bolometric corrections from Flower (1996) and \cite{Ber95} for Sirius B, 40 Eri B, and Procyon B. All relevant quantities are collected in Table~\ref{tbl-comp}. With much scatter the PNNi and WD appear to exhibit an approximate absolute bolometric magnitude-temperature relation, one that would hold, assuming similar radii and masses for all PNNi and WD. Both Equation 6 and the residuals, $\Delta$M$_{bol}$, argue otherwise.} 
\label{fig-6}
\end{figure}
\clearpage

\begin{figure}
\epsscale{1.00}
\plotone{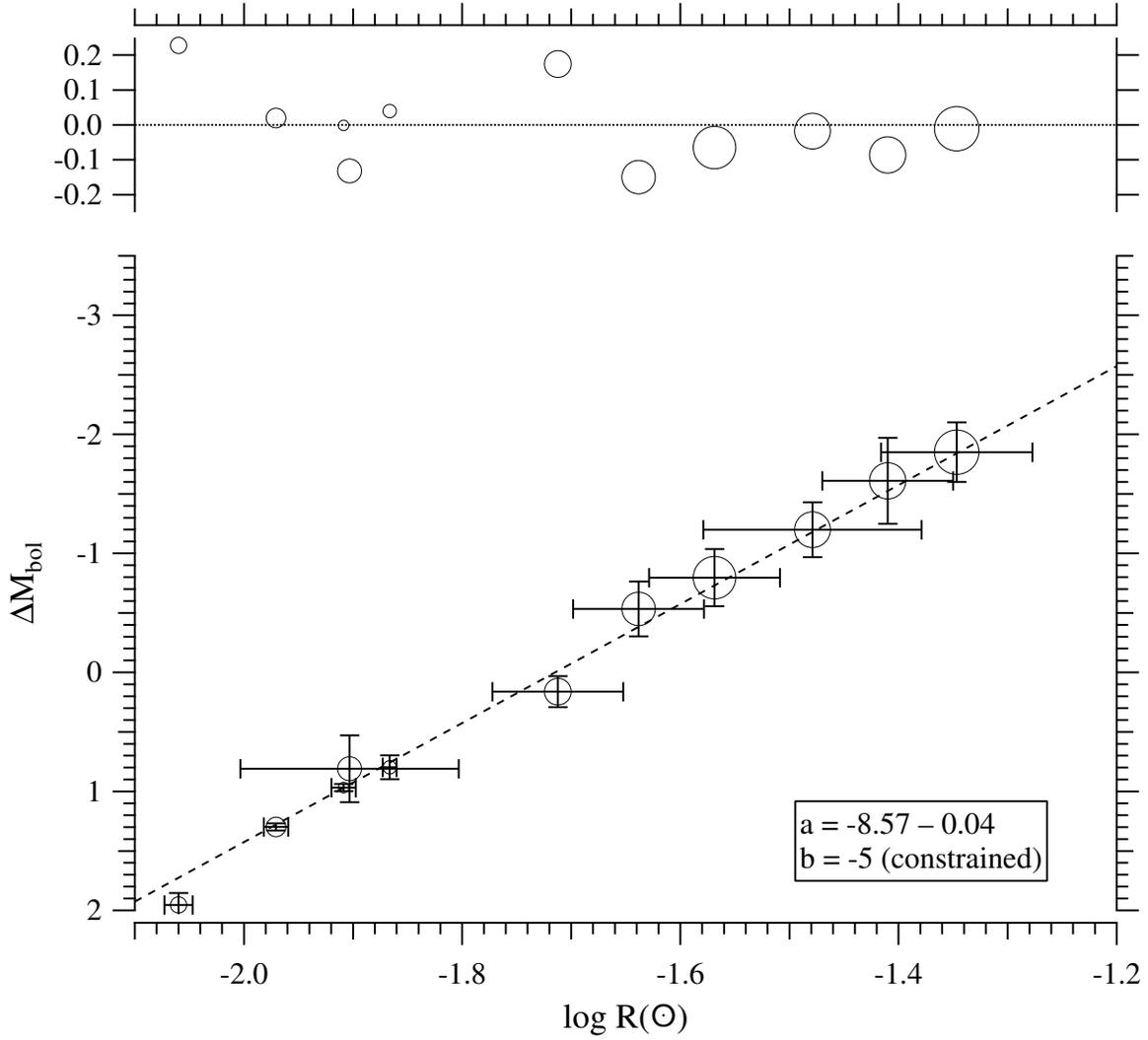}
\caption{Residuals,  $\Delta$M$_{bol}$, from the linear M$_{bol}$ - log T relationship for PNNi and WD in Figure~\ref{fig-6} plotted against the logarithm of the radius in solar units. The slope is constrained to the value expected from equation 6. The symbol size is proportional to surface temperature. The expected linear residual correlation (-~-~-) with log R exhibits an RMS scatter of 0.12 magnitudes.} 
\label{fig-7}
\end{figure}
\clearpage

\begin{figure}
\epsscale{0.75}
\plotone{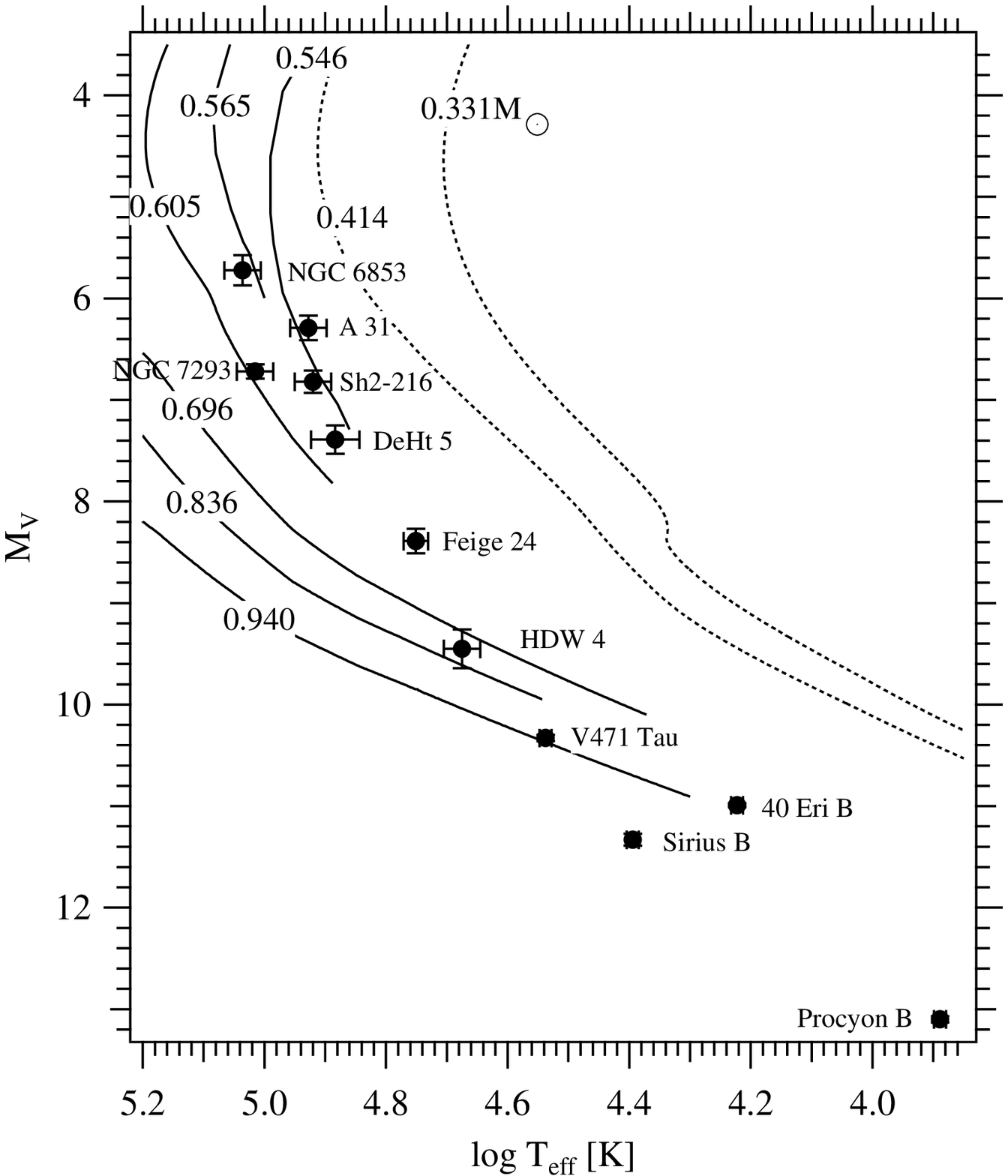}
\caption{PNNi and WD absolute magnitude, M$_V$ plotted against log temperature.  M$_V$ for the PNNi comes from our parallax, the apparent magnitude, and the interstellar absorption, A$_V$ (Table~\ref{tbl-AQ}) with two additional (HDW\,4, Sh2-216) from \cite{Har07}. The WD values are from \cite{Ben00a} for Feige 24, from \cite{OBr01} for V471 Tau, and calculated from Provencal (2002). All relevant quantities are collected in Table~\ref{tbl-comp}. The higher-mass (solid line) evolutionary tracks are from \cite{Sch96}. Lower-mass (dashed line) tracks are from  \cite{Dri99}.  The PNNi clump around M=0.57M$_{\sun}$.} 
\label{fig-Mvall}
\end{figure}
\clearpage

\begin{figure}
\epsscale{0.75}
\plotone{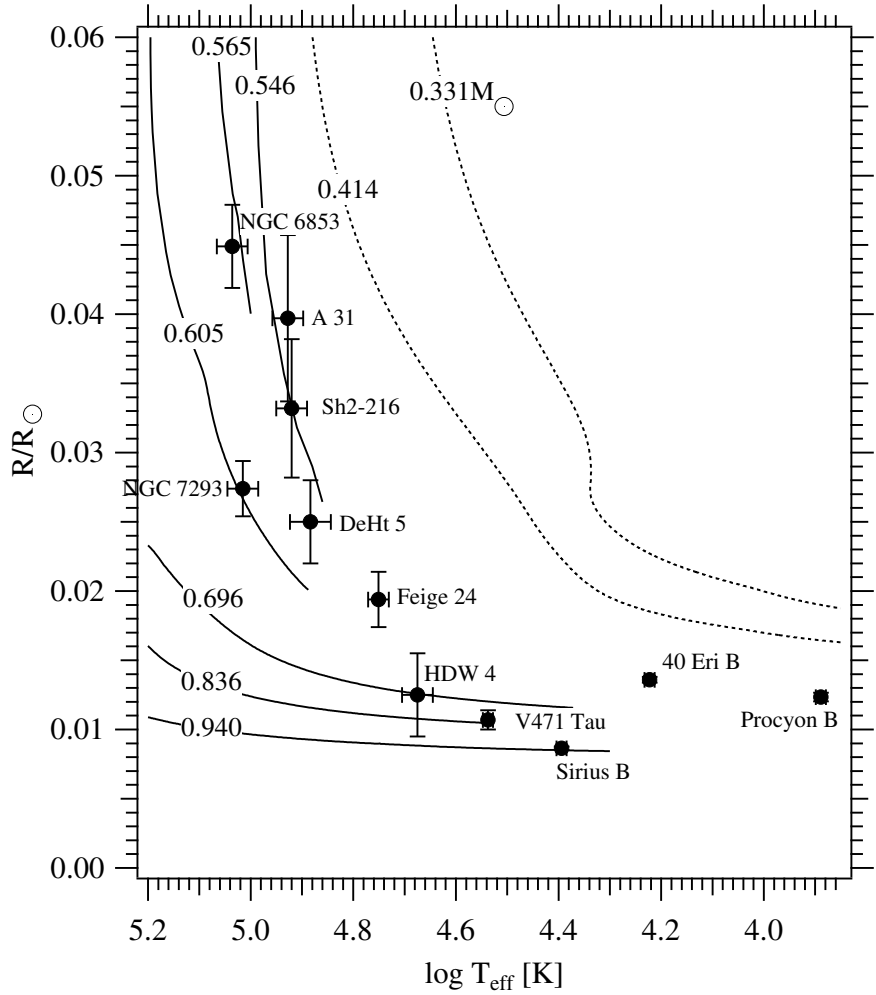}
\caption{PNNi and WD radii from Table~\ref{tbl-comp} plotted against log temperature. The evolutionary tracks for objects of high and low masses are from \cite{Sch96} and \cite{Dri99}. In this mapping the WD Procyon B and 40 Eri B appear to have masses consistent with previous measurements.}
\label{fig-Rall}
\end{figure}
\clearpage

\begin{figure}
\epsscale{1.00}
\plotone{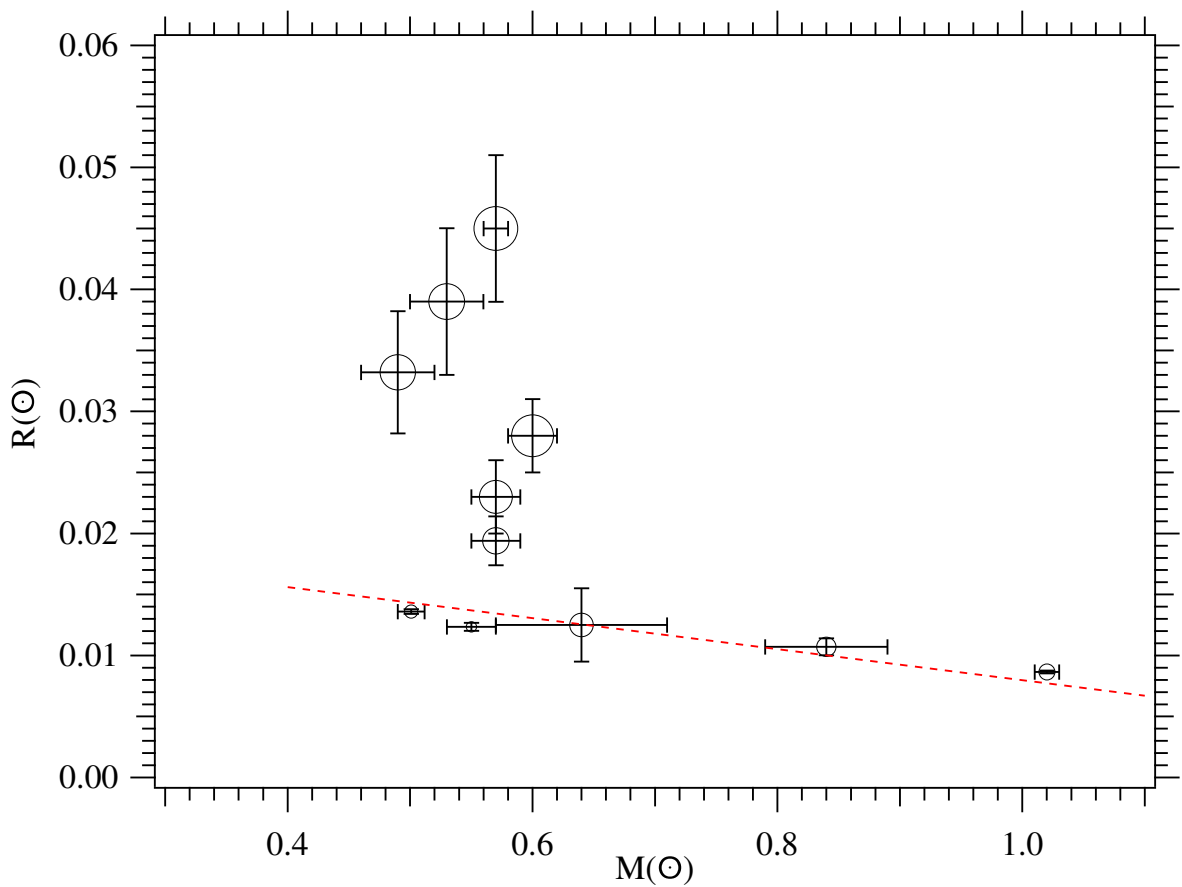}
\caption{Mass-Radius diagram for six PNNi and five WD with precise  radii and masses  (Table~\ref{tbl-comp}). From top to bottom we plot NGC\,6853, Abell 31, Sh2-216, NGC 7293, DeHt 5, Feige 24, 40 Eri B, HDW 4, Procyon B, V471 Tau, and Sirius B. Symbol size is proportional to surface temperature. The dashed line is the Hamada \& Salpeter (1961) carbon core relationship. Except for the coolest PNN, HDW 4, the PNNi (and the hottest WD, Feige 24) have larger radii than the cooler WD. } 
\label{fig-8}
\end{figure}

\begin{figure}
\epsscale{0.85}
\plotone{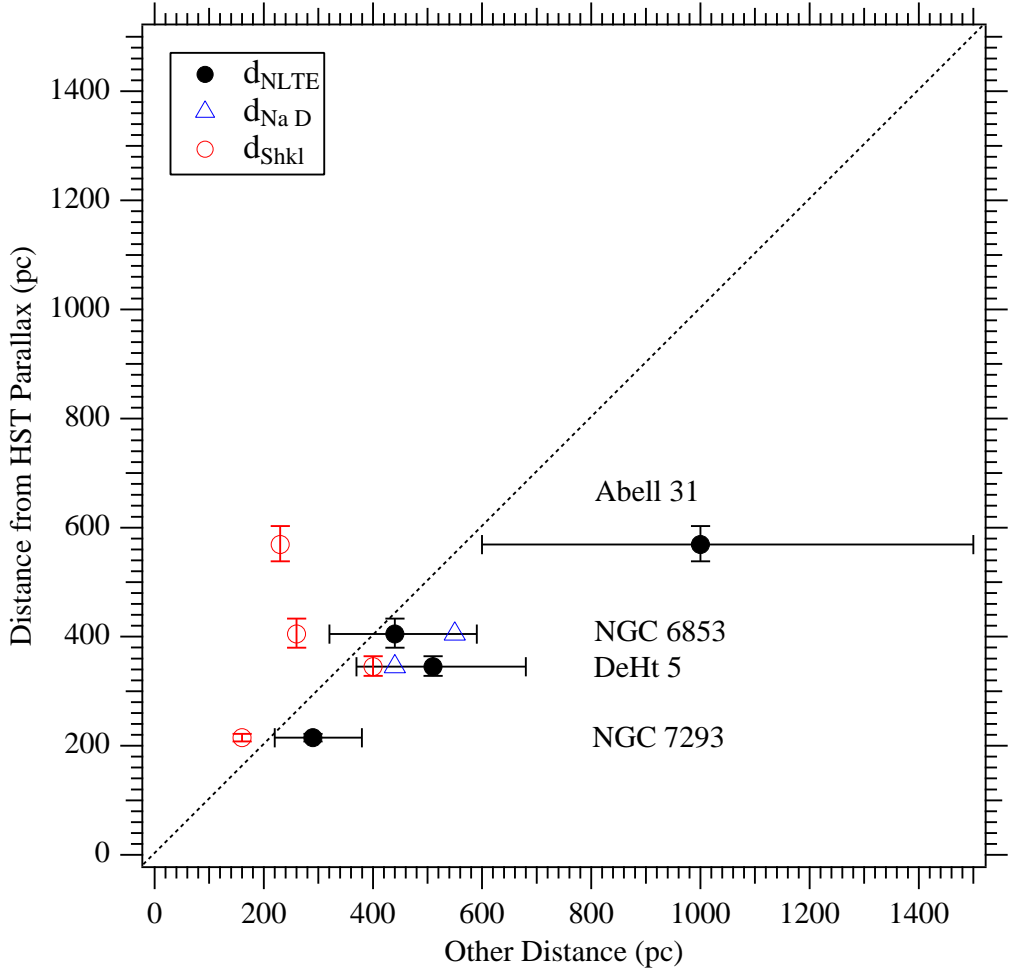}
\caption{Distances from weighted averages of \HST (this paper) and USNO (Harris et al. 2007) parallaxes compared to distances from NLTE analysis ($\bullet$Napiwotzki 2001), H$\beta$-derived distances from Shklovski (o 1956), and  distances estimated from interstellar Na D ($\triangle$ Napiwotzki \& Sch\"onberner 1995). The dashed line represents perfect agreement. Objects are labeled to the right or top.} 
\label{fig-dist}
\end{figure}
\clearpage

\end{document}